\documentclass[sigconf,nonacm]{acmart}
\AtBeginDocument{%
  }

\usepackage[linesnumbered,ruled,vlined]{algorithm2e}
\usepackage{algorithm2e}
\usepackage{bm}
\usepackage{subcaption}
\usepackage{hyperref}
\usepackage{soul}
\usepackage{xcolor}
\usepackage{tabularx}
\usepackage{multirow} 
\usepackage{array} 
\usepackage{rotating}
\usepackage{pdfpages}
\usepackage{url}
\usepackage{hyperref}

\usepackage{pifont}
\usepackage{threeparttable}
\usepackage{amsmath}
\usepackage{enumitem}
\usepackage{url}
\usepackage{xspace} 
\usepackage{listings, mdframed}
\usepackage{tcolorbox}

\newcommand{\name}{KubeGuard\xspace}

\begin{document}

\title{
\name:
LLM-Assisted Kubernetes Hardening via Configuration Files and Runtime Logs Analysis
}

\author{Omri Sgan Cohen, Ehud Malul, Yair Meidan, Dudu Mimran, Yuval Elovici, Asaf Shabtai}
\affiliation{%
  \institution{Department of Software and Information Systems Engineering \\ 
  Ben-Gurion University of The Negev}
   \country{Israel}
}


\begin{abstract}\label{abstract}
The widespread adoption of Kubernetes (K8s) for orchestrating cloud-native applications has introduced significant security challenges, such as misconfigured resources and overly permissive configurations.
Failing to address these issues can result in unauthorized access, privilege escalation, and lateral movement within clusters.
Most existing K8s security solutions focus on detecting misconfigurations, typically through static analysis or anomaly detection. 
In contrast, this paper presents \name, a novel runtime log-driven recommender framework aimed at mitigating risks by addressing overly permissive configurations.  
\name is designed to harden K8s environments through two complementary tasks: Resource Creation and Resource Refinement. 
It leverages large language models (LLMs) to analyze manifests and runtime logs reflecting actual system behavior, using modular prompt-chaining workflows. 
This approach enables \name to create least-privilege configurations for new resources and refine existing manifests to reduce the attack surface.
\name's output manifests are presented as recommendations that users (e.g., developers and operators) can review and adopt to enhance cluster security.
Our evaluation demonstrates that \name effectively generates and refines K8s manifests for Roles, NetworkPolicies, and Deployments, leveraging both proprietary and open-source LLMs. 
The high precision, recall, and F1-scores affirm \name's practicality as a framework that translates runtime observability into actionable, least-privilege configuration guidance. 
\end{abstract}

\maketitle

\section{Introduction}\label{sec:intro}

Kubernetes (K8s) is an open-source container orchestration platform that has revolutionized how cloud-native applications are deployed and managed. 
By automating container scheduling, scaling, and resource allocation, K8s enables highly efficient and resilient distributed infrastructures~\cite{kubernetes_overview}.
However, as the complexity of K8s clusters increases, securing them becomes challenging, particularly due to the prevalence of misconfigurations in K8s configuration files (manifests)~\cite{rahman2023security}.
These misconfigurations expose organizations to risks such as unauthorized access, lateral movement, and privilege escalation, underscoring the need for robust and adaptive security solutions that help users apply secure configurations and reduce the likelihood of human error during policy definition~\cite{kampa2024navigating}.

Most K8s security solutions rely on predefined configurations~\cite{shamim2020xi,shamim2021mitigating}, static  analysis~\cite{samir2023adaptive,mahajan2022detection,diarra2024depth}, or industry-standard rule-based  tools~\cite{checkov,terrascan,kics,kubelinter}. 
Basic hardening can be achieved using native K8s controls, for example, replacing default allow-all settings with tailored NetworkPolicies or refining overly permissive RBAC Roles; however, both depend on precise configuration and are therefore challenging to maintain at scale.
Moreover, although these methods effectively reduce the attack surface, they often lack visibility into runtime behavior and struggle to adapt to dynamic environments and evolving threats.
Even advanced approaches leveraging artificial intelligence (AI) and large language models (LLMs)~\cite{malul2024genkubesec,li2405llm,ye2025llmsecconfig} typically exhibit limited responsiveness to real-time system states.
Several studies have employed runtime logs for monitoring and anomaly detection in K8s~\cite{kitahara2020highly,li2024ai,sarika2023automating,cao2022learning}. 
More recent methods leveraged various logs using AI, but fed them into static pipelines~\cite{ye2025llmsecconfig} or generated configurations for just a single resource type (e.g., only NetworkPolicies)~\cite{kimkubeteus}, often not even K8s-native~\cite{khan2025lads}.

\begin{figure}[b]
    \centering
    \includegraphics[width=1\linewidth, trim=20 10 21 5, clip]{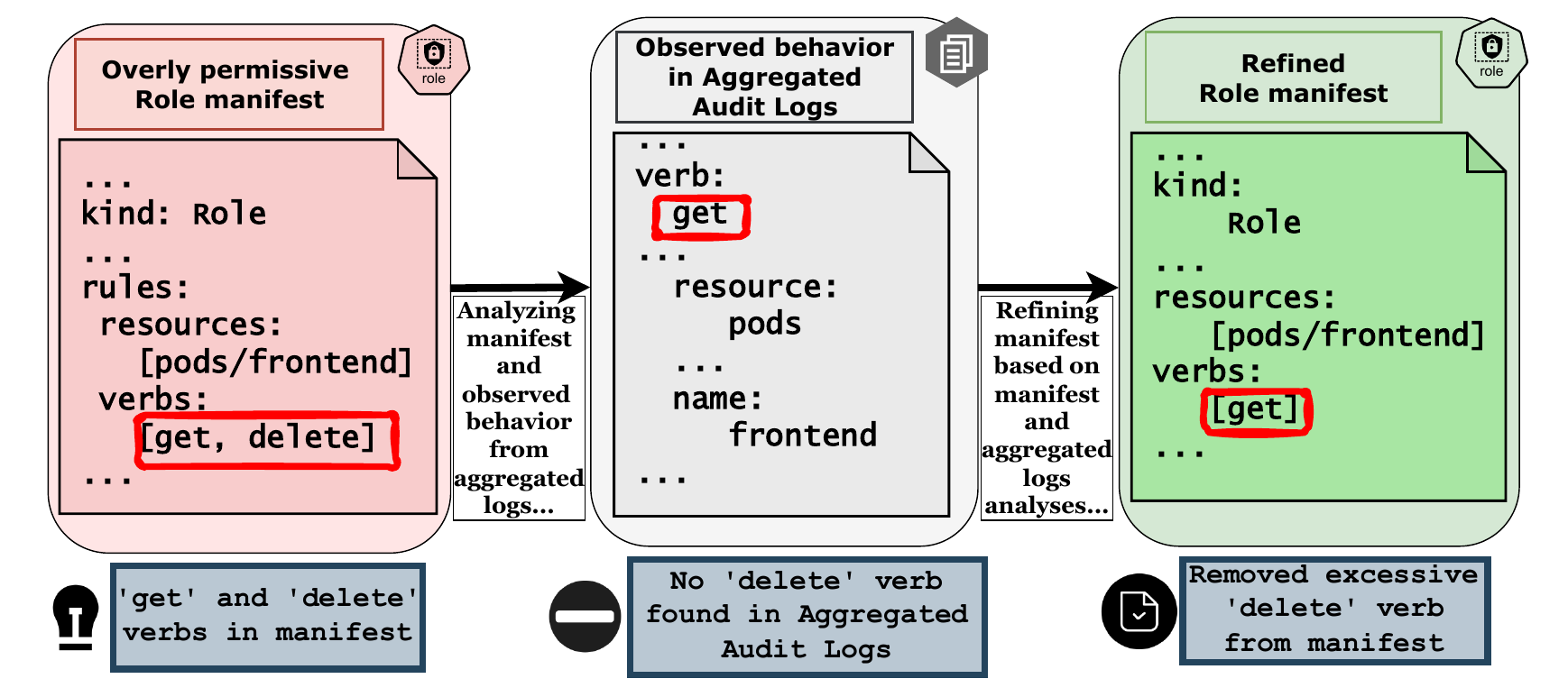}
    \caption{Example of log-driven manifest hardening. 
    An overly permissive K8s Role is refined by aligning the manifest with observed runtime behavior in aggregated logs.}
    \label{fig:hardening_example}
\end{figure}

To address the aforementioned limitations of existing approaches, keep pace with the demands of dynamic environments, and support the required shift from reactive diagnostics to proactive defense, this paper introduces \emph{\name}: a novel LLM-based framework for dynamically hardening K8s systems based on prompt-chaining workflows~\cite{anthropic_chain_prompts} that leverage multi-source runtime logs (a log-driven hardening example is provided in Fig.~\ref{fig:hardening_example}). 
\name uses audit logs~\cite{k8s_auditing}, provenance data~\cite{gehani2012spade}, and network traffic flows~\cite{hubble_observability} to extract actionable insights and systematically create or refine K8s manifests. 
These heterogeneous logs are aggregated into structured activity records, allowing \name to correlate user actions, network activity, and resource-level changes. 
The aggregation preserves behavioral security context while compressing high-volume logs into concise summaries (reducing token count by $99.96\%$), enabling efficient LLM analysis within token limits.

\name acts as a security-aware recommender framework that assists developers and system owners in hardening K8s environments by suggesting least-privilege configurations based on observed application behavior. 
It addresses two complementary tasks, both aimed at reducing the attack surface of K8s-based systems: \emph{Resource Creation} and \emph{Resource Refinement}.
Resource Creation focuses on generating secure log-derived manifests for RBAC Roles and NetworkPolicies, directly from observed application behavior.
Additionally, Resource Refinement hardens existing manifests by 
suggesting elimination and restriction of excessive permissions and unnecessary rules in Roles, NetworkPolicies, and Deployments, based on aggregated runtime logs. 
By integrating structured prompts into modular chain workflows, \name aligns security configurations with application behavior and best practices, ensuring robust and adaptive protection throughout the cluster lifecycle.

We evaluated \name using two microservice-based applications~\cite{google_microservices_demo,azure_store_demo}, each comprising multiple manifests. 
Log data was collected from audit~\cite{k8s_auditing}, provenance~\cite{gehani2012spade}, and network traffic~\cite{hubble_observability} sources during different operational phases.
Following an aggregation step, these logs served as input for resource creation and refinement tasks. 
As part of our preliminary evaluation, we examined several prompting strategies, including zero-shot~\cite{syed2025zero}, chain-of-thought (CoT)~\cite{wei2022chain}, and prompt chaining~\cite{anthropic_chain_prompts}, and evaluated multiple proprietary LLMs to identify the most effective configuration for \name. 
Prompt chaining outperformed the other techniques and was adopted as our core approach, and GPT-4o~\cite{openai_gpt4o_system_card} achieved the highest accuracy and generalization, therefore selected as the representative proprietary LLM for further evaluation.

To support \name's broader adoption, we extended our evaluation from commercial LLMs to open-source small language models (SLMs).
Therefore, our comprehensive evaluation addresses the needs of two organizational profiles: those comfortable with external data sharing via API, typically associated with commercial LLMs, and those requiring stricter privacy and having limited computational resources~\cite{kibriya2024privacy,wang2025comprehensivesurveyllmagentstack}.
We evaluated multiple SLMs using prompt chaining and selected Llama-3.1-8B~\cite{grattafiori2024llama} as the most effective local model, balancing performance and deployability.

Evaluation results prove \name's ability to generate valid, hardened K8s manifests for various tasks and model sizes. 
Using GPT-4o, \name achieved a perfect F1-score in Role Creation ($1.00$) and high F1-scores in Role ($0.953$), NetworkPolicy ($0.961$), and Deployment ($0.929$) refinements, demonstrating prompt chaining effectiveness. 
Although less accurate, Llama-3.1-8B still performed well, reaching an F1-score of $0.808$ in Role Refinement and $0.793$ in NetworkPolicy Creation, affirming it as a practical privacy-preserving alternative.

The primary contributions of this research are as follows:
\begin{itemize}[nosep, leftmargin=*]
    \item \textbf{A comprehensive framework for hardening K8s resources, supporting both resource creation and refinement:}
    By both creating and refining K8s resources, \name enables broad applicability to diverse workloads and security postures, allowing users to enforce least-privilege principles in both newly developed and already deployed K8s environments. 
    In addition, unlike previously proposed solutions that are limited to one specific resource type~\cite{bringhenti2023security,kimkubeteus}, \name operates across multiple critical K8s resources: Roles, NetworkPolicies, and Deployments.
    
    \item \textbf{Log-driven and context-aware configuration optimization for K8s:}
    \name adopts a log-centric approach, using runtime telemetry to inform configuration decisions.
    Unlike static tools~\cite{checkov,terrascan,kics,kubelinter}, it analyzes multi-source logs, including audit, network, and provenance, collected from the running environment.
    This enables context-aware configuration generation and refinement from application behavior, adapting to changes over time and reducing the attack surface empirically and dynamically.

    \item \textbf{Model-agnostic prompt chaining using LLMs and SLMs:}
    \name integrates large and small language models for context-aware reasoning on logs and manifests, using multiple structured prompt-chaining pipelines. 
    This design supports both high-accuracy API models
    and privacy-preserving local models, providing flexibility across different organizational constraints.
    
\end{itemize}

\begingroup\footnotesize
\begin{table*}[h]
\centering
\caption{Comparative overview of related work on K8s hardening.}\label{tab:related_work}
\begin{tabularx}{\textwidth}{|X|>{\raggedright\arraybackslash}X|>{\raggedright\arraybackslash}X|>{\raggedright\arraybackslash}X|>{\raggedright\arraybackslash}X|
}
\hline
\textbf{Paper}
& \textbf{Method}             
& \textbf{Input}                                  
& \textbf{Goal}
& \textbf{Affected Resources}
\\ 
\hline
KGSecConfig~\cite{haque2022kgsecconfig}    
    & Knowledge graphs-based & Official docs, security reports  & Misconfiguration detection and correction 
    & Container orchestrators configurations 
    \\
    \hline
GenKubeSec~\cite{malul2024genkubesec}     
    & Fine-tuned and prompted LLMs & K8s manifests             & Misconfiguration detection and correction 
    & K8s manifests 
    \\
    \hline
LLMSecConfig~\cite{ye2025llmsecconfig}    
    & Retrieval-Augmented Generation, LLM prompting   & Static analysis tools output, source code, docs & Misconfiguration correction 
    & Container-focused K8s manifests 
    \\
    \hline
Don't Train, Just Prompt~\cite{kratzke2024don} 
    & Zero- and few-shot prompting, prompt chaining & User-described desired K8s Deployments & K8s resources creation 
    & K8s manifests 
    \\
    \hline
Kub-Sec~\cite{zhu2022kub}                  
    & Rule-based        & Audit logs                             & AppArmor profile generation 
    & AppArmor policies for Pods 
    \\
    \hline
LADs~\cite{khan2025lads}                  
    & Retrieval-Augmented Generation, few-shot and CoT prompting, prompt chaining  & Natural language user intent, failure logs, docs, configuration examples & Cloud configurations creation and refinement 
    & Configurations of K8s-deployed cloud services 
    \\
    \hline
KubeFence~\cite{cesarano2025kubefencesecurityhardeningkubernetes} 
    & Rule-based & Helm charts, K8s API requests & K8s API security policies generation 
    & K8s security policies 
    \\
\hline
KUBETEUS~\cite{kimkubeteus}               
    & Fine-tuned and prompted LLMs & Natural language intents, real-time K8s API data, configuration files & CNI-compliant K8s NetworkPolicies creation 
    & K8s NetworkPolicies 
    \\
    \hline
\textbf{Our framework (\name)}
    & Prompt chaining & K8s manifests, audit, provenance, and network logs & K8s resources creation and refinement 
    & K8s Roles, NetworkPolicies, and Deployments 
    \\
\hline
\end{tabularx}
\end{table*}
\endgroup

\section{Related Work}\label{sec:rel_work}

We summarize prior work on K8s hardening in Table~\ref{tab:related_work}, describing their methods, inputs, goals, and the resources they affect.

\subsection{K8s Misconfiguration Detection and Mitigation
}

Mitigating misconfigurations in K8s has been widely studied, highlighting risks such as overprovisioned permissions and misconfigured RBAC Roles ~\cite{shamim2020xi,kampa2024navigating}. 
Broader cloud-native practices, aimed at securing Function-as-a-Service~\cite{prechtl2020investigating} and DevOps workflows~\cite{alawneh2022expanding,myceksecurity}, offer insight but do not address K8s specifics. 
Most tools emphasize rule-based static analysis~\cite{checkov,terrascan,kubelinter,kics} and vulnerability assessment and detection~\cite{rahman2023security,kubehunter,hirschberg2022kubescape,morfonios2023kubernetes,rangta2022tools}.
Complementary work includes misconfiguration detection during deployment~\cite{mahajan2022detection} and manifest verification tools~\cite{diarra2024depth}. 
Although valuable, these efforts mostly detect deviations from best practices without adaptive remediation and rely on predefined rules that overlook runtime context.

Moving beyond static checks, advanced runtime-capable approaches have emerged. 
An adaptive controller diagnosed misconfigurations during execution~\cite{samir2023adaptive}, and KGSecConfig synthesized secure manifests using knowledge graphs, given high-quality input~\cite{haque2022kgsecconfig}. 
Recent AI-driven approaches embedded LLMs throughout misconfiguration detection, localization, reasoning, and remediation ~\cite{malul2024genkubesec}, or together with static analysis for vulnerability detection and mitigation~\cite{li2405llm,ye2025llmsecconfig}, or used LLMs for real-time anomaly detection in K8s~\cite{sheng2024research,diaf2024bartpredict}. 
However, these methods typically do not adapt configurations according to observed runtime behavior.

Logs have been widely used for monitoring and anomaly or failure analysis, leveraging machine learning~\cite{koryugin2023analysing,mustyala2022comprehensive,kitahara2020highly,li2024ai,sarika2023automating,cao2022learning}, 
and also for misconfiguration reporting~\cite{russell2024centralized}, provenance-driven methods~\cite{abbas2022paced,hassan2018towards}, and network-based risk prediction~\cite{ruan2021network}. 

However, these works rarely translate insights into concrete hardening actions. 
In contrast, \name leverages multi-source runtime logs to create and refine K8s manifests, shifting from detection of misconfigurations to adaptive, behavior-aware generation of least-privilege configurations as workloads evolve.

\subsection{Secure K8s Configurations Generation}\label{subsec:secure_k8s_config_generation}

Beyond static modeling, AI and LLM-based methods have emerged to automate K8s security configurations, including risk profiling with operational data context~\cite{schreiber2025ai}, workload migration~\cite{ueno2024migrating}, generating Ansible playbooks for remediation~\cite{ansible,sarda2024leveraging}, and translating natural language into manifests~\cite{kratzke2024don}. 
However, these approaches overlook K8s-specific resources and often lack runtime context, risking misconfigurations or overly permissive defaults~\cite{malul2024genkubesec}.

Building on runtime integration, Kub-Sec used audit logs to generate AppArmor profiles~\cite{apparmor,zhu2022kub}.
LADs applied Retrieval-Augmented Generation ~\cite{lewis2020retrieval}, few-shot learning, and iterative feedback~\cite{madaan2023self} to optimize settings from deployment failures~\cite{khan2025lads}, though reactively.

Several works focused on generating network-layer configurations, mainly through automatic NetworkPolicy synthesis~\cite{bringhenti2023security,npguard}.
In addition, KUBETEUS proposed an intention-driven framework with fine-tuned LLMs to generate NetworkPolicies~\cite{kimkubeteus}. 
Beyond the network layer, KubeFence~\cite{cesarano2025kubefencesecurityhardeningkubernetes} derived API request filters from trusted Operator manifests~\cite{kubernetes_operator_pattern} and audit2rbac~\cite{liggitt_audit2rbac_2023} demonstrated log-driven K8s RBAC generation.
An empirical comparison of \name to audit2rbac and KUBETEUS is presented in Sec.~\ref{subsec:Empirical_Comparison_to_Existing_Methods}.
Although effective in scope, these methods remain limited to network or audit configurations and lack integration of diverse runtime data. 

In contrast, \name combines audit, network, and provenance logs to provide a holistic security view, creating and refining manifests for Roles, NetworkPolicies, and Deployments.

\begin{figure*}[h]
    \centering
    \includegraphics[width=1.00\linewidth, trim=10 10 10 10, clip]{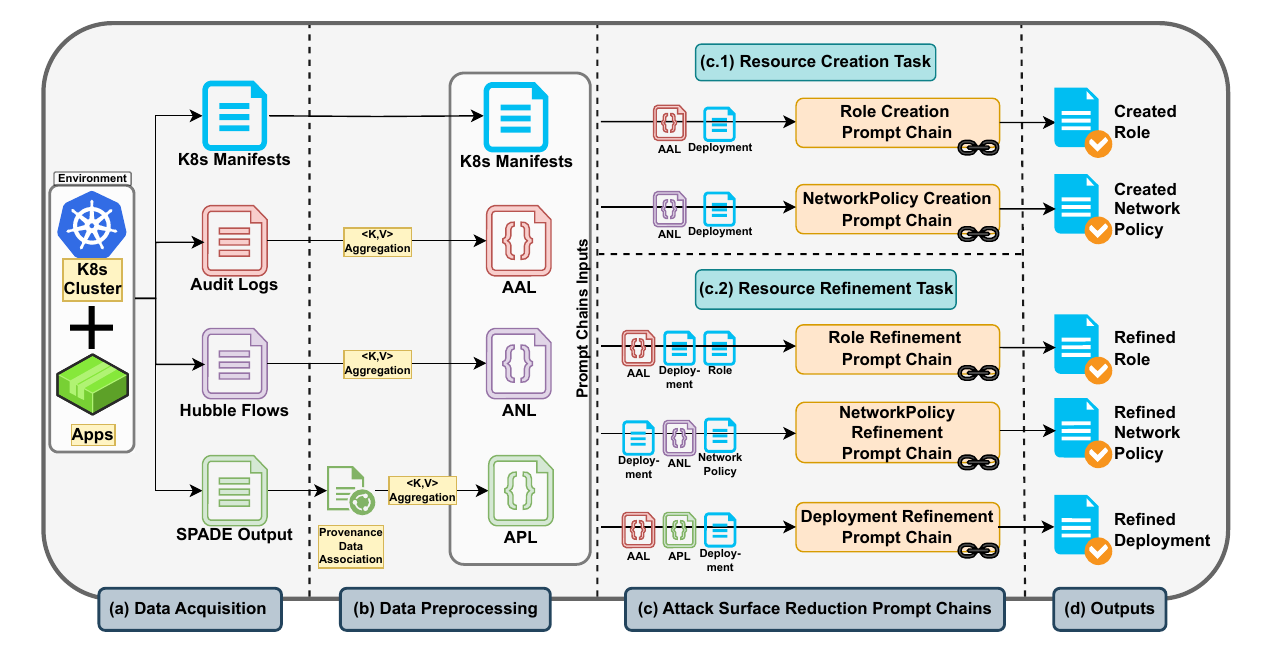}
    \caption{
    Overview of the \name framework: (a) collect manifests and runtime logs, (b) preprocess into aggregated behavioral datasets, (c) apply LLM-based prompt chains for attack surface reduction, and (d) output least-privilege manifests.
    }
    \label{fig:proposed_method}
\end{figure*}

\section{Proposed Method}\label{sec:Proposed_Method}

Our proposed framework, \name, illustrated in Fig.~\ref{fig:proposed_method}, adaptively enhances K8s security configurations by analyzing runtime logs.
It begins by preprocessing data from multiple sources and then employs a language model using task-tailored prompt chains.
Prompt chaining involves using the output of one prompt as the input of the next, facilitating multistep reasoning~\cite{wu2022ai}.
Through this approach, \name systematically progresses from data analysis to manifest optimization, creating or refining core K8s resources: Deployments~\cite{k8s_deployment}, Roles~\cite{k8s_role}, and NetworkPolicies~\cite{k8s_network_policies}.

\name is designed as an assistive recommender framework.
System operators can selectively invoke desired prompt chains and apply the output manifests across standard application lifecycle environments. 
The hardened manifests generated by \name are correctly formatted and ready to be applied.
\name does not automatically enforce them, so operators can review and adopt recommended configurations while aligning with compliance requirements and operational best practices.
The LLM-based reasoning provided along with the generated manifests gives users additional transparency and control to apply hardening actions confidently, based on actual runtime behavior.
\name, described in detail below, 
upholds the least-privilege principle and strengthens operational security, all while preserving application functionality.

\subsection{Data Acquisition and Preprocessing}\label{subsec:data_acquisition_and_preprocessing}
The first part of the proposed method consists of two stages: ``Data Acquisition'' (Fig.~\ref{fig:proposed_method}(a)), in which logs are collected from the K8s environment, and ``Data Preprocessing'' (Fig.~\ref{fig:proposed_method}(b)), where logs are aggregated into structured context-rich inputs for LLM analysis.

\paragraph{Data Acquisition}
K8s manifests define the configuration of workloads and resources within a cluster, therefore serving as a baseline for further optimization (example manifest provided in Listing~\ref{lst:nginx_manifest}).
Given a task (creating or refining a K8s resource), \name receives logs as input, from three primary data sources, in addition to the manifest (as illustrated in Fig.~\ref{fig:proposed_method}(a)):
(1) K8s audit logs~\cite{k8s_auditing}, which capture K8s metadata operations, 
(2) Hubble flows~\cite{hubble_observability}, which detail network communication patterns, and 
(3) SPADE output~\cite{gehani2012spade} that depicts system-level interactions.
These log types are used to provide a complementary view of cluster behavior; audit logs record API requests and authorization decisions, making them indispensable for inferring RBAC permissions; Hubble flows capture pod-level ingress and egress traffic, enabling precise alignment of NetworkPolicies with actual communication patterns; and SPADE’s provenance data records low-level OS events within containers, linking operations back to microservices and providing crucial container runtime information.
All logs are collected using dedicated open-source tools: K8s auditing~\cite{k8s_auditing} for audit logs, SPADE~\cite{SPADE} for provenance, and Hubble~\cite{hubble_github} for network traffic.

\begin{figure}[h]
    \centering
    \includegraphics[width=0.8\linewidth, trim=20 20 20 20, clip]{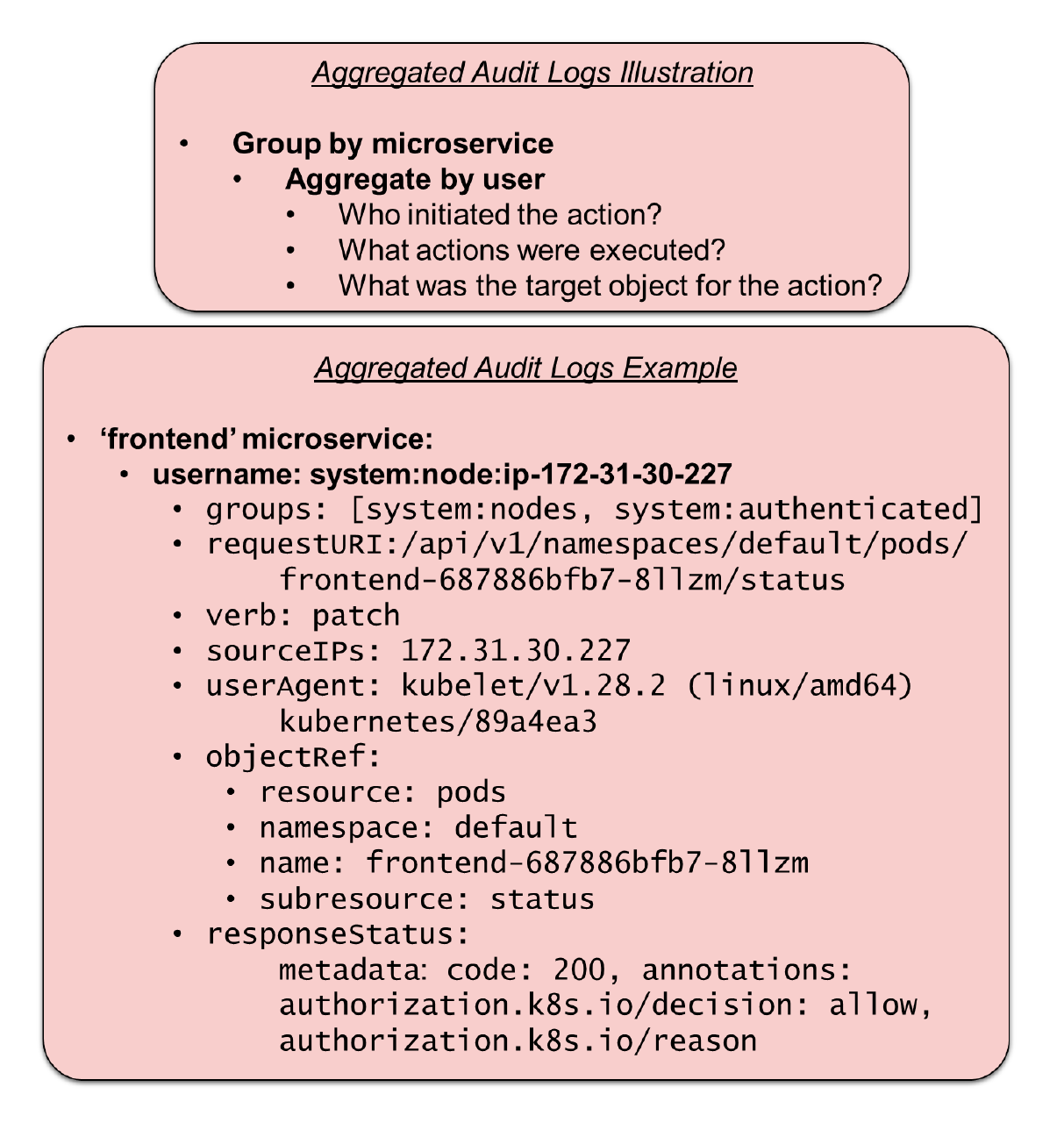}
    \caption{Illustration and example of Aggregated Audit Logs (AALs). The keys of the AALs are of the original audit logs, with their values aggregated by microservice and user.}
    \label{fig:AAL_example}
\end{figure}

\paragraph{Data Preprocessing}
Addressing LLM token and context constraints, \name employs a structured log aggregation approach that condenses raw logs while preserving all essential information (illustrated as ``\texttt{<K,V> Aggregation}'' in Fig.~\ref{fig:proposed_method}(b) and exemplified in Fig.~\ref{fig:AAL_example}). 
Each field (key) in the logs is used as a map key, and its corresponding values are recursively extracted and aggregated into a \texttt{key→set of values} (K–V) HashMap. 
Not only does this improve organization and readability, but it also minimizes log complexity without compromising critical details (details provided in Algorithm~\ref{alg:log_aggregation}). 
Notably, K-V aggregation reduces log size by up to $99.96\%$ in model input tokens, making large-scale analysis feasible.
To maintain context and coherence, logs are aggregated by microservices and logical entities (e.g., users or provenance components~\cite{moreau2011open}).
This also enables injecting relevant contextual information along the aggregated logs (SPADE provenance data model~\cite{spade_audit_provenance} and Hubble predefined network endpoints identities~\cite{hubble_special_identities}), helping LLMs better interpret the structure and semantics of the condensed logs.
Thus, K-V aggregation transforms raw logs into concise input for efficient and accurate LLM analysis within computational limits.

The following outlines each \name input, including their data acquisition and preprocessing steps.

\subsubsection{K8s Manifest Files}\label{subsubsec:K8s_Manifest_Files}
YAML-formatted specifications that define the desired state of resources within a K8s cluster~\cite{kubernetes_manifest}. 
These files contain essential configuration details for managing K8s components such as Pods, Roles, and NetworkPolicies (a Deployment manifest example provided in Listing~\ref{lst:nginx_manifest}). 
K8s manifest files are converted into JSON format for usability and uniformity.

\subsubsection{Aggregated Audit Logs (AALs)}\label{subsubsec:Aggregated_Audit_Logs}
Structured datasets derived from K8s audit logs, summarizing security-relevant operations within a cluster.
The raw logs are collected through K8s auditing~\cite {k8s_auditing}, following a minimal audit policy that logs all API requests at the \texttt{Metadata} level~\cite{kubernetes_minimal_audit_policy}.
They are then grouped by microservices and aggregated through K-V aggregation, as illustrated and exemplified in Fig.~\ref{fig:AAL_example}.
AALs consolidate details such as request origins and authorization decisions into RBAC-relevant summaries that serve as the ground truth for exercised API actions, enabling inference of least-privilege Roles and removal of unused permissions.

\subsubsection{Aggregated Network Logs (ANLs)}\label{subsubsec:Aggregated_Network_Logs}
Structured datasets derived from Hubble output, a network observability tool that captures real-time Pod-level communication events within a K8s cluster~\cite{hubble_observability,hubble_github}.
Hubble network flows are aggregated by Pods using K-V aggregation, resulting in a structured representation of interservice communications (illustration and example provided in Fig.~\ref{fig:ANL_example}). 
Although provenance data can represent network activity through system calls, it does not natively relate to K8s resources (see Sec.~\ref{subsubsec:Aggregated_Provenance_Logs}).
In contrast, Hubble-originated ANLs enrich network flows with K8s metadata (e.g., Pods, IPs, ports, traffic directions) and directly map higher-layer network events to workloads.
This enables immediate correlation of network activity with Pods, facilitating precise and efficient optimization of K8s NetworkPolicies.

\subsubsection{Aggregated Provenance Logs (APLs)}\label{subsubsec:Aggregated_Provenance_Logs}

Structured provenance datasets, derived from SPADE output~\cite{gehani2012spade,SPADE}.
APLs document system-level interactions within a K8s cluster by representing causal relationships among components, thus enabling interpretation and correlation of system behavior (example provided in Fig.~\ref{fig:APL_example}).
Provenance complements audit and network data by revealing in-container runtime information outside the K8s control plane visibility (e.g., process executions, file writes, socket activity), allowing identification of workload behaviors relevant to security hardening.

Next, we outline the complete preprocessing of SPADE output leading to its integration into \name as APLs.

\begin{figure}[h]
    \centering
    \includegraphics[width=\linewidth, trim=20 5 20 5, clip]{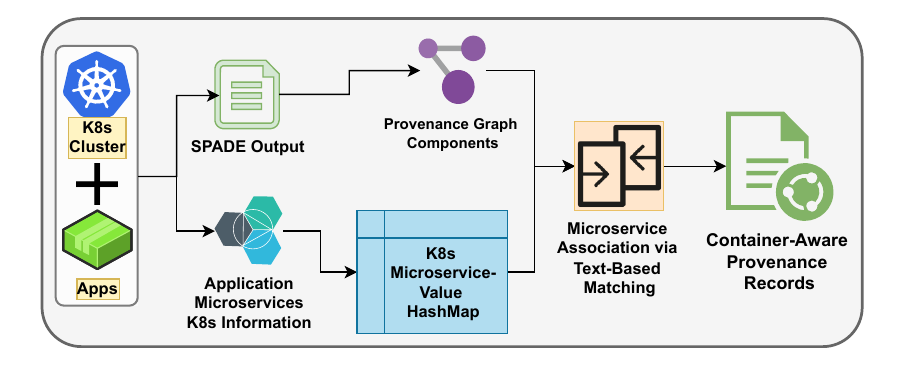}
    \caption{Illustration of the provenance data association process. 
    SPADE output is matched with K8s microservices information to produce container-aware provenance records.}
    \label{fig:prov_association}
\end{figure}

First, we collect raw provenance logs using SPADE and its extension CLARION~\cite{chen2021clarion,CLARION}, which provides Linux namespace awareness~\cite{linux_namespaces} for containerized environments (e.g., K8s).
Then, these logs undergo preprocessing, referred to as ``Provenance Data Association'', as seen in Fig.~\ref{fig:proposed_method}(b) and depicted in Fig.~\ref{fig:prov_association}.
In this process, we transform SPADE output, including its container context provided by CLARION, into structured records of provenance graph (PG) edges (events) and vertices (artifacts or processes)~\cite{gehani2012spade,moreau2011open}.
To associate PG nodes and edges with their corresponding application microservices, we construct a \texttt{microservice→value} HashMap with microservice information from the K8s cluster (labels, ports, etc.).
We then perform text-based matching between this information and the PG components' attributes, for example, aligning process command lines or executable paths with container images or binaries, or mapping socket IP/port pairs with corresponding K8s Service/Pod metadata.
This association step also serves as a filter, as PG components that do not map to any known microservice are discarded, reducing downstream volume.
The result is container-aware provenance records enriched with K8s microservices context, ready for further processing.
Subsequently, K-V aggregation is applied to these container-aware records by events, producing an APLs dataset representing high-level system behaviors. 

\subsection{Attack Surface Reduction Prompt Chains}\label{subsec:Attack_Surface_Reduction_Prompt_Chains}

\begin{figure*}[]
    \centering
    \includegraphics[width=1.00\textwidth, trim=15 15 20 5, clip]{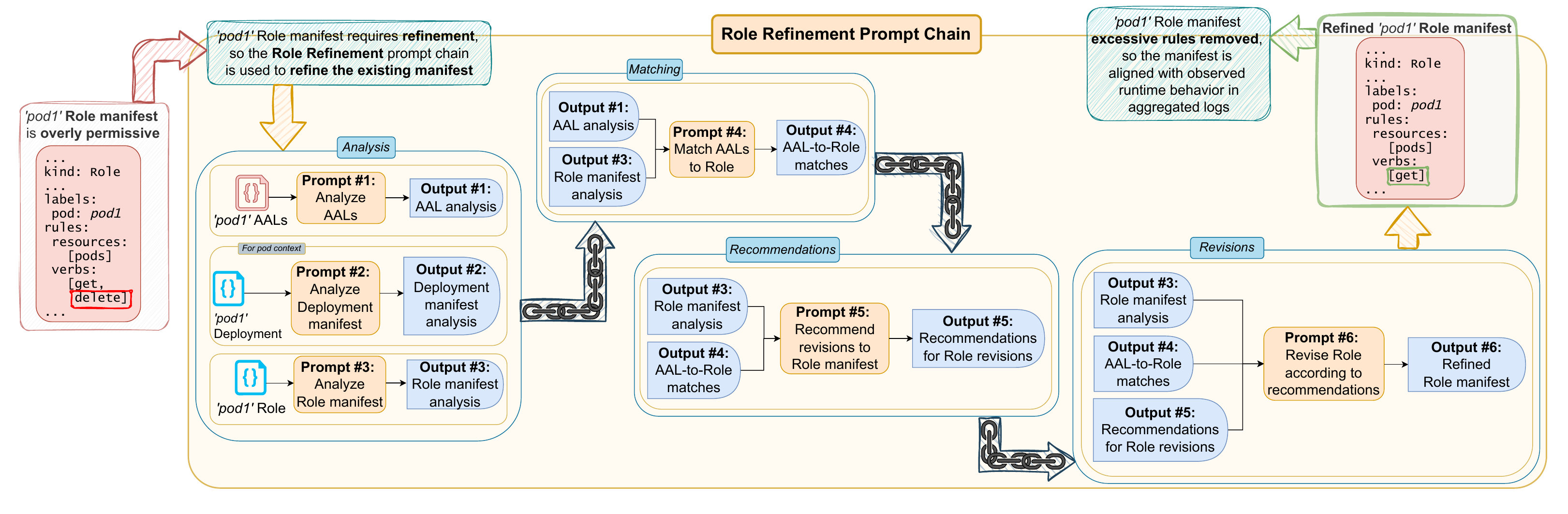}
    \caption{Illustration of the Role Refinement prompt chain process: 
    An overly permissive Role is analyzed, matched with aggregated audit logs (AALs), revised according to recommendations, and output as a hardened Role.
    }
    \label{fig:role_refinement_running_example}
\end{figure*}

As shown in Fig.\ref{fig:proposed_method}(c), our framework performs two complementary tasks, handling both missing and existing configurations, as follows.
\begin{enumerate}[nosep, leftmargin=*]
    \item \textbf{Resource Creation:}
    When an application is missing a required configuration (e.g., absent Role or NetworkPolicy), \name creates a new, secure manifest from observed runtime behavior, enforcing least-privilege access control, which hardens K8s security (example of a creation prompt chain provided in Fig.~\ref{fig:networkpolicy_creation_prompt_chain}).
    
    \item \textbf{Resource Refinement:}
    When a manifest already exists but is overly permissive, \name suggests stricter configurations according to runtime logs, generating a hardened manifest by the least-privilege principle, while maintaining application functionality (refinement prompt chain illustrated in Fig.~\ref{fig:role_refinement_running_example}).
\end{enumerate}

These two tasks are implemented through five structured prompt chains, invoked at users’ discretion, each tailored for a target K8s resource and a main task, integrating relevant AALs, ANLs, and APLs. 
The prompt chains output new manifests as recommendations rather than directly modifying the cluster, so generated manifests can be reviewed and selectively applied to align with organizational policies and application needs.
All prompts across all chains follow a carefully designed reasoning structure comprising fixed elements: role (system prompt), task, requirements, instructions, input, and expected output with accompanying justification~\cite{ucd2024promptstructure}, while allowing modularity to support diverse workflows.

The following are descriptions of the prompt chains developed:

\subsubsection{Resource 
Creation}\label{subsubsec:Resource_Creation}
This task covers the generation of new, secure K8s manifests through two prompt chains: \emph{Role Creation} and \emph{NetworkPolicy Creation} (Fig.~\ref{fig:proposed_method}(c.1)).
The logical sequencing of these chains begins with analyses of manifests and aggregated logs, followed by prompts that generate the required resources.\\

\textbf{Role Creation:}\label{bullet:Role_Creation} 
The Role Creation prompt chain generates K8s ServiceAccounts~\cite{kubernetes_service_accounts}, Roles, and RoleBindings~\cite{k8s_role} from Deployment manifests and AALs, aligning RBAC policies with workload requirements to enforce fine-grained access control under the least-privilege principle.
The prompt chain involves (1) analyzing AALs to derive observed RBAC context, (2) extracting security insights and vulnerabilities from the Deployment manifest, (3) creating scoped ServiceAccounts, (4) generating Roles with minimal required permissions, and (5) binding Roles to ServiceAccounts by defining RoleBindings.
This logical sequence ensures each step's insights guide the generation of subsequent resources.
The output is a secure, minimal Role manifest embedded within a cohesive RBAC model, including ServiceAccounts, Roles, and RoleBindings, to ensure consistency with K8s logic and seamless cluster integration.\\

\textbf{NetworkPolicy Creation:}\label{bullet:networkpolicy_creation} 
This prompt chain generates K8s NetworkPolicies to enforce Pod-level ingress and egress traffic control by aligning Deployment manifests with observed patterns from ANLs (illustrated in Fig.~\ref{fig:networkpolicy_creation_prompt_chain}).
NetworkPolicy Creation consists of (1) analyzing the Deployment manifest to extract security-relevant network configurations, (2) examining ANLs to detect communication patterns, anomalies, and unauthorized access, and (3) generating NetworkPolicies with fine-grained ingress and egress traffic rules.
The resulting NetworkPolicy manifest ensures secure inter-Pod communication and mitigates potential vulnerabilities.

\subsubsection{Resource Refinement}\label{subsubsec:Resource_refinement}
This task hardens K8s security by refining overly permissive manifests based on log analysis, through three prompt chains: \emph{Role Refinement}, \emph{NetworkPolicy Refinement}, and \emph{Deployment Refinement} (Fig.~\ref{fig:proposed_method}(c.2)).
The chains' sequencing logic is based on first analyzing manifests and aggregated logs, matching their findings, and then generating optimization recommendations before revising the target resource manifest accordingly.\\

\textbf{Role Refinement:}\label{bullet:Role_refinement}
This prompt chain optimizes existing K8s Roles by eliminating unnecessary permissions and enforcing least-privileged access, as illustrated in Fig.~\ref{fig:role_refinement_running_example}. 
It (1) analyzes AALs to derive security-relevant RBAC insights, (2) evaluates the Deployment manifest for security risks, (3) examines the Role manifest to detect excessive rules and verbs, (4) matches Role definitions with AALs to assess actual usage, (5) generates permission-reduction recommendations, and (6) revises the Role accordingly. 
The output Role manifest ensures minimal RBAC allocation to harden security, reducing the attack surface while maintaining operational integrity.\\
    
\textbf{NetworkPolicy Refinement:}\label{bullet:NetworkPolicy_refinement}
This prompt chain refines existing K8s NetworkPolicies by enforcing stricter ingress and egress controls and communication rules based on ANLs.
The prompt chain (1) analyzes the Deployment manifest for network-relevant configurations, (2) evaluates ANLs for observed traffic patterns, (3) examines the existing NetworkPolicy manifest for security gaps, (4) matches the NetworkPolicy with observed network flows from ANLs to identify misconfigurations and (4.A) validates these log-to-policy matches, (5) generates recommendations for refinement, and (6) revises the NetworkPolicy accordingly.
The resulting NetworkPolicy manifest ensures more precise and secure traffic control, aligning with K8s' additive-restrictive NetworkPolicy logic.\\

\textbf{Deployment Refinement:}\label{bullet:Deployment_refinement}
This prompt chain refines existing K8s Deployments.
It analyzes AALs, APLs, and the manifest itself and reduces excessive permissions and misconfigurations (illustrated in Fig.~\ref{fig:high_level_deployment_refinement}).
The prompt chain (1) extracts RBAC insights from AALs,
(2) analyzes APLs to identify only necessary dependencies, 
(3) assesses the Deployment manifest for vulnerabilities,
(4) matches the Deployment analysis against the AALs' analysis and (5) the APLs' analysis to detect security-relevant patterns,
(6) generates actionable manifest recommendations,
and (7) revises the Deployment manifest to incorporate the recommended changes.
The output is a refined Deployment manifest, ensuring reduced overprovisioning while maintaining application functionality.

\section{Experimental Setup}\label{sec:experimental_setup}

To quantitatively evaluate \name, we used two microservice applications and collected their audit, provenance, and network logs. 
Using both synthetic and real manifests, we evaluated multiple prompting methods and model categories across tasks, including Role, NetworkPolicy, and Deployment creation and refinement. 
Our experiments also cover prompt strategy design, model benchmarking, comparison to other methods, sensitivity testing, and ablation studies, using precision, recall, and F1-score as key metrics.

\subsection{Environment Setup}\label{subsec:env_setup}

\subsubsection{LLMs}\label{subsubsec:llm_env}
We accessed proprietary LLMs via API and local open-source SLMs through Hugging Face~\cite{wolf2019huggingface}.
All local model evaluations were conducted on machines with NVIDIA RTX 6000 GPU~\cite{rtx6000}.

\subsubsection{K8s Environment}\label{subsubsec:k8s_env}

We deployed two K8s-based applications in a self-managed cluster on AWS infrastructure: 
\textit{Google's Microservices Demo} (Online Boutique)~\cite{google_microservices_demo} 
and \textit{AKS Store Demo}~\cite{azure_store_demo}. 
Their logs were collected across standard deployment environments, as representative traffic was produced using the applications’ native load generators to simulate realistic usage and ensure comprehensive endpoint coverage~\cite{microservicesdemo_loadgenerator,aksstoredemo_virtualcustomer,aksstoredemo_virtualworker}.
Open-source tools were used to collect log data.
Audit logs were recorded at the \texttt{Metadata} level via K8s' auditing mechanism~\cite{k8s_auditing}, using a minimal audit policy~\cite{kubernetes_minimal_audit_policy}. 
Network traffic data was captured using the Hubble observability tool~\cite{hubble_github}. 
Provenance data was collected using SPADE~\cite{gehani2012spade,SPADE} and its CLARION extension for containerized environments~\cite{chen2021clarion,CLARION}.
This setup is reproducible with the same workloads and tools.

\subsection{Task-Specific Evaluation}\label{subsec:data_and_tasks}

\name's tasks were evaluated through log-based assessments by comparing their output manifests with least-privilege baseline policies, derived from real application logs.
Refinement tasks were also evaluated using manifests with synthetically injected test cases (see Sec.~\ref{subsec:synthetic_manifests}).
For Resource Creation, \name was evaluated on generating secure, accurate, and complete manifests based on behavioral log data. 
For Resource Refinement, \name was evaluated on limiting over-permissive manifests using runtime behavior observed in the logs.

\subsubsection{Synthetic Manifests with Excessive Permissions}\label{subsec:synthetic_manifests}

\begin{figure}[h]
    \centering
    \includegraphics[width=1\linewidth, trim=20 15 20 15, clip]{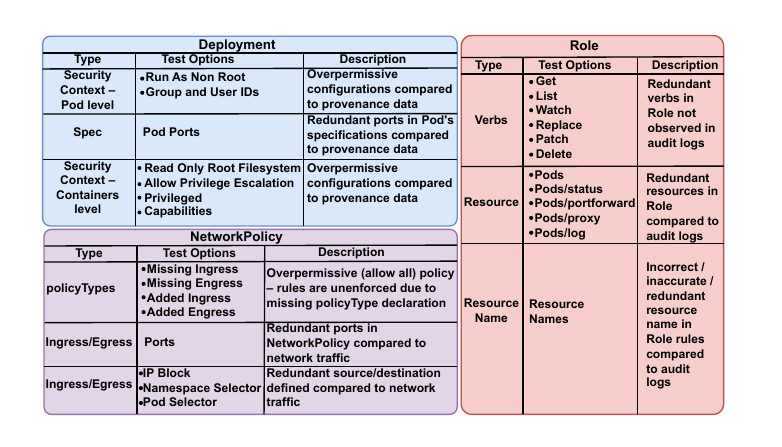}
    \caption{Rule taxonomy used to define and evaluate excessive or insecure configurations across core K8s resources.
    }
    \label{fig:rule_book}
\end{figure}

To evaluate refinement in a controlled and measurable manner, we designed synthetic K8s manifests containing security anti-patterns, either by intentionally injecting them or by preserving misconfigurations already present in the original files. 
These excessive manifests span the three resource types: Roles, NetworkPolicies, and Deployments.
The design of these manifests is guided by the taxonomy presented in Fig.~\ref{fig:rule_book}, which includes misconfigurations such as redundant permissions, overly permissive rules, and unused resource specifications—typical issues observed in real-world scenarios.
By using excessive configurations according to this taxonomy, we also enable quantifiable measurement of the attack surface reduction by \name, directly linking hardening results to taxonomy rules.

\subsection{Evaluation Metrics}\label{subsec:eval_metrics}

\subsubsection{Task Effectiveness}\label{subsubsec:Task_Effectiveness}

To evaluate \name's effectiveness across Resource Creation and Resource Refinement tasks, we adopted standard classification metrics: precision, recall, and F1-score (F1).
These metrics are based on the following definitions:

\noindent\textbf{True Positive (TP):}
A permission, rule, or configuration, correctly generated or refined based on runtime behavior observed in logs.

\noindent\textbf{False Positive (FP):} 
A permission, rule, or configuration added or retained incorrectly without evidence from observed behavior (e.g., an extra Role verb or an incorrect NetworkPolicy rule port).

\noindent\textbf{False Negative (FN):}
A required permission or rule that is missing from the hardened output, either because it was not created or not refined by the framework, despite being evidenced in the logs.

\noindent\textbf{True Negative (TN):}
A configuration element that was correctly retained, as it did not require hardening.

\subsubsection{Similarity Among Aggregated Logs}\label{subsubsec:Similarity_among_Aggregated_Logs}

We also assessed how closely aggregated logs from different data collection durations resemble each other using three bag‑of‑tokens scores, which anchor the log-duration sensitivity results discussed in Sec.~\ref{subsubsec:data_logs_sens}:

\noindent\textbf{Cosine Similarity:} $\text{Cosine}(A, B) = \frac{\sum_{i=1}^{n} A_i B_i}{\sqrt{\sum_{i=1}^{n} A_i^2} \cdot \sqrt{\sum_{i=1}^{n} B_i^2}}$~\\
Similarity between two token-frequency vectors via the cosine of the angle between them, capturing how consistently tokens appear regardless of their total counts.

\noindent\textbf{Overlap Coefficient:} $\text{Overlap}(A, B) = \frac{|A \cap B|}{\min(|A|, |B|)}$~\\
Fraction of shared tokens in the smaller set, highlighting whether all tokens from the shorter window appear in the longer one.

\noindent\textbf{Dice Coefficient (Sørensen–Dice):} $\text{Dice}(A, B) = \frac{2 \cdot |A \cap B|}{|A| + |B|}$~\\
Twice the shared tokens divided by the total tokens in both sets, balancing overlap with overall size to reward shared content and penalize mismatches.

\subsection{Experimental Design}\label{subsec:exp_phases}

In our experiments, we examined prompting strategies, model categories, task-specific performance, and robustness.
We began by benchmarking three prompting techniques (zero-shot~\cite{syed2025zero}, CoT~\cite{wei2022chain}, and prompt chaining~\cite{anthropic_chain_prompts}) using several proprietary language models applied to a representative task: NetworkPolicy Creation (described in Sec.~\ref{bullet:networkpolicy_creation} and Fig.~\ref{fig:networkpolicy_creation_prompt_chain}).
This phase aimed to identify the most effective prompting strategy and select the best-performing model for subsequent experiments.
Once the prompting approach and LLM were selected, we leveraged their combination across all supported tasks: creating Roles and NetworkPolicies, and refining existing manifests for Deployments, Roles, and NetworkPolicies.

For local open-source SLMs, we first evaluated various models on the same representative task used for LLM selection.
Then, we selected the best-performing SLM for complete evaluation of all tasks.
To improve SLM output, we introduced an iterative refinement mechanism that recursively applies the refinement prompt chain to harden the manifest over multiple rounds.
To investigate the effect of model scale, we evaluated a larger local LLM on selected tasks, measuring its ability to surpass SLMs. 

We also empirically compared \name with two other relevant methods.
In addition, we conducted a sensitivity analysis to study the impact of prompt chain ordering and log collection duration on \name's performance. 
Ablation studies isolated the contributions of specific inputs and prompt chain components (e.g., data explanations, matching steps, log types) to \name's performance. 
Error analysis was integrated throughout the results section (Sec.~\ref{sec:results}) to examine performance limitations.

\section{Results}\label{sec:results}

This section presents the experimental results of \name on our testbed applications in multiple evaluation phases.
We first identified the best-performing LLM and prompting strategy: GPT-4o and prompt chaining.
We then extended their evaluation to multiple tasks and K8s resources.
We also addressed privacy and computational constraints by evaluating local SLMs, selecting Llama-3.1-8B as the most capable open-source option. 
We repeated the complete task evaluation using this model.
Recognizing the value of even marginal reductions of the attack surface, we explored recursive prompt chaining, showing that iterative refinement can incrementally enhance security.
Additionally, we investigated whether larger local models yield better results.
We also empirically compared \name's capabilities with two other task-equivalent methods.
Lastly, sensitivity analysis revealed how prompt order and log collection duration affect performance, while ablation studies quantified the individual contributions of prompt components.

\subsection{Selecting the Best LLM and Prompting Approach}\label{subsec:selecting_the_best_llm_and_prompting_approach}

Our first experiment was designed to identify the most effective LLM and prompting technique.
To streamline this resource-intensive evaluation (five prompt chains, three models, three techniques), we selected NetworkPolicy Creation as a balanced baseline—complex enough to challenge models, yet requiring relatively short prompting. 
We evaluated three prompting techniques: zero-shot~\cite{syed2025zero} (baseline), CoT~\cite{wei2022chain}, and prompt chaining~\cite{anthropic_chain_prompts}, across three proprietary LLMs: GPT-4o~\cite{achiam2023gpt}, Gemini-1.5-Pro~\cite{pichai2023introducing}, and Claude-3.5-Sonnet~\cite{claude}.

The results, summarized in Table~\ref{tab:best-llm-and-prompting-comparison}, demonstrate how different prompting strategies impacted model performance:
In the zero-shot setting (prompt example provided in Fig.~\ref{fig:zero_shot_example}), GPT-4o achieved the highest precision ($0.819$) 
and Claude led in recall ($0.830$) and F1 ($0.764$), 
while Gemini trailed behind (F1 of $0.700$). 
This setting revealed limited task comprehension across all models.
In comparison, structured reasoning through CoT prompting~\cite{wei2022chain} showed notable performance gains (prompt example provided in Fig.~\ref{fig:cot_example}). 
Gemini outperformed other models in all metrics, scoring an F1 of $0.827$, while GPT-4o and Claude remained competitive with F1 of $0.775$ and $0.767$, respectively; however, GPT-4o had a more consistent performance.
Prompt chaining delivered the strongest results (prompt chain example provided in Fig.~\ref{fig:networkpolicy_creation_prompt_chain}).
GPT-4o achieved the highest performance on all metrics: precision $0.945$, recall $0.926$, and F1 $0.935$, outperforming Claude and Gemini.
As seen in Table~\ref{tab:best-llm-and-prompting-comparison}, although Gemini performed best with CoT, prompt chaining produced the highest F1-scores for the other two models and the best overall performance, justifying its selection as our core technique.
We hypothesize that this advantage stems from the sequential nature of our approach, which aligns well with the stepwise structure of prompt chaining. 
Notably, although GPT-4o did not outperform all models in every setting, it consistently showed competitive results across strategies and excelled in prompt chaining, affirming its selection as the primary model for subsequent evaluations.

\begin{table}[]
\centering
\caption{Performance comparison of proprietary LLMs on NetworkPolicy Creation across prompting techniques.}
\resizebox{\columnwidth}{!}{
\begin{tabular}{|l|l|c|c|c|c|}
\hline
\textbf{Technique} 
& \textbf{Metric} 
& \textbf{GPT-4o} 
& \textbf{Gemini-1.5-Pro} 
& \textbf{Claude-3.5-Sonnet} \\
\hline
\multirow{4}{*}{Zero-shot} 
    & Precision 
    & \bm{$0.819 \pm 0.090$}
    & $0.758 \pm 0.154$ 
    & $0.737 \pm 0.193$ \\
    & Recall    
    & $0.706 \pm 0.108$ 
    & $0.685 \pm 0.175$ 
    & \bm{$0.830 \pm 0.165$} \\
    & F1-Score
    & $0.750 \pm 0.072$ 
    & $0.700 \pm 0.015$ 
    & \bm{$0.764 \pm 0.154$} \\
\hline
\multirow{4}{*}{CoT} 
    & Precision 
    & $0.817 \pm 0.109$ 
    & \bm{$0.854 \pm 0.111$} 
    & $0.820 \pm 0.173$ \\
    & Recall    
    & $0.753 \pm 0.127$ 
    & \bm{$0.822 \pm 0.147$} 
    & $0.751 \pm 0.125$ \\
    & F1-Score  
    & $0.775 \pm 0.097$ 
    & \bm{$0.827 \pm 0.102$} 
    & $0.767 \pm 0.108$ \\
\hline
\multirow{4}{*}{
Chaining} 
    & Precision 
    & \bm{$0.945 \pm 0.079$} 
    & $0.869 \pm 0.136$ 
    & $0.924 \pm 0.063$ \\
    & Recall    
    & \bm{$0.926 \pm 0.107$} 
    & $0.732 \pm 0.207$ 
    & $0.752 \pm 0.060$ \\
    & F1-Score  
    & \bm{$0.935 \pm 0.093$} 
    & $0.784 \pm 0.207$ 
    & $0.827 \pm 0.047$ \\
\hline
\end{tabular}
}
\label{tab:best-llm-and-prompting-comparison}
\end{table}

\subsection{Evaluating the Best Proprietary LLM: GPT-4o}\label{subsec:evaluating_the_best_proprietary_LLM_GPT4o}

Following the selection of GPT-4o as the top-performing proprietary LLM (Sec.~\ref{subsec:selecting_the_best_llm_and_prompting_approach}), we evaluated it on all five prompt chains (Sec.~\ref{subsec:Attack_Surface_Reduction_Prompt_Chains}).
The results are summarized in Table~\ref{tab:gpt4o-performance} and discussed below.

\begin{table}[h]
\centering
\caption{
Performance of GPT-4o across resource creation and refinement prompt chains.
}
\resizebox{\columnwidth}{!}{
\begin{tabular}{|l|l|l|l|}
\hline
\textbf{Task} & \textbf{Precision} & \textbf{Recall} & \textbf{F1-Score} \\ \hline
Role Creation 
& $1.00 \pm 0.0$ & $1.00 \pm 0.0$ & $1.00 \pm 0.0$ \\ \hline
NetworkPolicy Creation 
& $0.945 \pm 0.079$ & $0.926 \pm 0.107$ & $0.935 \pm 0.093$ \\ \hline
Role Refinement 
& $1.00 \pm 0.0$ & $0.914 \pm 0.084$ & $0.953 \pm 0.047$ \\ \hline
NetworkPolicy Refinement 
& $0.990 \pm 0.019$ & $0.940 \pm 0.113$ & $0.961 \pm 0.068$ \\ \hline
Deployment Refinement 
& $0.975 \pm 0.053$ & $0.894 \pm 0.140$ & $0.929 \pm 0.100$ \\ \hline
\end{tabular}
}
\label{tab:gpt4o-performance}
\end{table}

\noindent\textbf{Role Creation:}~\label{par:gpt4o_role_creation}
GPT-4o achieved perfect scores across all metrics ($1.00$), demonstrating its ability to generate accurate and complete Role definitions from log-derived behavioral insights, without introducing unnecessary permissions or missing required ones.
\noindent\textbf{NetworkPolicy Creation:}~\label{par:gpt4o_networkPolicy_creation}
The model achieved high precision ($0.945$) and recall ($0.926$), resulting in F1 of $0.935$. 
These results show that GPT-4o reliably identifies the necessary communication policies. 
A minor drop in recall was observed due to occasional underspecification of ingress or egress rules.
This issue can be mitigated through postprocessing; however, it also increases the risk of overfitting.
\noindent\textbf{Role Refinement:}~\label{par:gpt4o_role_refinement}
GPT-4o maintained perfect precision and achieved high recall ($0.914$) and F1 ($0.953$). 
These results highlight the model’s strong capacity to refine RBACs using audit data.
Lower recall was observed in infrastructure-facing services (e.g., MongoDB~\cite{mongodb}, RabbitMQ~\cite{rabbitmq}), 
where additional access permissions not present in the logs were mistakenly retained. 
Alternatively, Roles can be created from the ground up using Role Creation, as it achieved perfect scores across all metrics.
\noindent\textbf{NetworkPolicy Refinement:}~\label{par:gpt4o_networkPolicy_refinement}
GPT-4o achieved near-perfect precision ($0.990$), strong recall ($0.940$), and F1 of $0.961$.
Error analysis showed that recall declined for services with high network traffic, where complex logs led the model to miss some communication patterns.
This limitation can be mitigated by postprocessing.
\noindent\textbf{Deployment Refinement:}~\label{par:gpt4o_deployment_refinement}
Using AALs and APLs, GPT-4o achieved precision of $0.975$, recall of $0.894$, and F1 of $0.929$, reflecting strong capabilities in hardening existing overly permissive Deployment manifests.
Despite the high precision, a small portion of FPs remained.
We investigated them and discovered that most were harmless modifications, mainly involving resource allocation.
While these refinements potentially improve security, they fall outside the scope of \name and were therefore treated as FPs in our evaluation.
Recall remained strong at $0.894$, although occasional omissions, such as incomplete port restrictions, resulted in a few missed refinement cases.

\subsection{Selecting the Best Local (Open-Source) SLM}\label{subsec:selecting_best_local_os_slms}

To address privacy and computational constraints associated with proprietary LLMs, particularly sending data outside of the organization due to external API usage~\cite{kibriya2024privacy}, we conducted experiments using three local open-source SLMs: 
Qwen2.5-7B~\cite{qwen2}, 
Ministral-8B~\cite{ministral}, 
and Llama-3.1-8B~\cite{grattafiori2024llama}. 
Each SLM was evaluated on the NetworkPolicy Creation baseline task, using prompt chaining, which was previously identified as the most effective strategy (Sec.~\ref{subsec:selecting_the_best_llm_and_prompting_approach}), and selected as our primary prompting technique.
This design choice enabled us to focus on model comparison.

\begin{table}[h]
\centering
\caption{
Performance comparison of local open-source SLMs on NetworkPolicy Creation using prompt chaining.
}
\resizebox{\columnwidth}{!}{
\begin{tabular}{|l|l|l|l|}
\hline
\textbf{Model} & \textbf{Precision} & \textbf{Recall} & \textbf{F1-Score}  \\ \hline
Qwen2.5-7B 
& $0.772 \pm 0.122$ & $0.763 \pm 0.193$ & $0.759 \pm 0.149$  \\ \hline
Ministral-8B
& $0.784 \pm 0.239$ & $0.692 \pm 0.271$ & $0.724 \pm 0.255$  \\ \hline
Llama-3.1-8B 
& \bm{$0.813 \pm 0.113$} & \bm{$0.803 \pm 0.253$} & \bm{$0.793 \pm 0.181$}  \\ \hline
\end{tabular}
}
\label{tab:experimenting-slms-netpol-creation}
\end{table}

As shown in Table~\ref{tab:experimenting-slms-netpol-creation}, Llama achieved the highest results;  precision of $0.813$, recall of $0.803$, and F1 of $0.793$. 
Qwen followed with a balanced performance, achieving a recall of $0.763$ and an F1 of $0.759$, though its precision ($0.722$) trailed slightly behind. 
Ministral showed competitive precision ($0.784$), outperforming Qwen in that metric, but had the lowest recall ($0.692$), resulting in the lowest F1 ($0.724$). 
These findings confirm that open-source SLMs can handle complex policy optimization tasks.
They are ideal for organizations with cost or privacy constraints, offering a balance between performance and accessibility.
While proprietary models lead in performance (Sec.~\ref{subsec:selecting_the_best_llm_and_prompting_approach}), it is important to note that even minor improvements in manifest security are valuable, as they reduce the attack surface by enforcing the principle of least privilege. 
Thus, SLMs offer a practical privacy-preserving alternative for secure on-premise deployments.
Building on the results outlined above, Llama was selected for subsequent experiments, reflecting prioritization of consistent and accurate outputs in security-sensitive contexts.

\subsection{Evaluating the Best Local SLM: Llama-3.1-8B}
\label{subsec:evaluating_the_best_local_SLM_Llama}

With Llama-3.1-8B identified as the best-performing SLM (Sec.~\ref{subsec:selecting_best_local_os_slms}), we replicated the 
GPT-4o experiments (Sec.~\ref{subsec:evaluating_the_best_proprietary_LLM_GPT4o}), covering all resource creation and refinement tasks (experiment results in Table~\ref{tab:llama-performance}). 

\begin{table}[h]
\centering
\caption{Performance of Llama-3.1-8B across resource creation and refinement prompt chains.}
\resizebox{\columnwidth}{!}{
\begin{tabular}{|l|l|l|l|}
\hline
\textbf{Task} & \textbf{Precision} & \textbf{Recall} & \textbf{F1-Score}  \\ \hline
Role Creation 
& $0.616 \pm 0.327$ & $0.689 \pm 0.244$ & $0.607 \pm 0.266$ \\ \hline
NetworkPolicy Creation 
& $0.813 \pm 0.113$ & $0.803 \pm 0.253$ & $0.793 \pm 0.181$ \\ \hline
Role Refinement 
& $0.975 \pm 0.065$ & $0.758 \pm 0.230$ & $0.808 \pm 0.197$ \\ \hline
NetworkPolicy Refinement 
& $0.889 \pm 0.191$ & $0.385 \pm 0.222$ & $0.504 \pm 0.228$ \\ \hline
Deployment Refinement 
& $0.689 \pm 0.159$ & $0.804 \pm 0.108$ & $0.728 \pm 0.089$ \\ \hline
\end{tabular}
}
\label{tab:llama-performance}
\end{table}

\noindent\textbf{Role Creation:}~\label{par:llama_role_creation}
Llama demonstrated modest effectiveness in generating K8s Roles, achieving a precision of $0.616$, a recall of $0.689$, and an F1 of $0.607$.
The model tended to generate more rules than were strictly necessary.
Although some of these additions improve security, they were treated as FPs since they were not evidenced in the logs, which contributed to reduced precision.
This behavior reflects a recurring pattern also observed with GPT-4o (Sec.~\ref{subsec:evaluating_the_best_proprietary_LLM_GPT4o}), where security-enhancing outputs occasionally extended beyond the defined task scope.
\noindent\textbf{NetworkPolicy Creation:}~\label{par:llama_networkPolicy_creation}
Llama exhibited strong performance for this task: precision of $0.813$, recall of $0.803$, and an F1 of $0.793$. 
The model effectively inferred required network rules from ANLs, producing secure and minimal policies. 
While the model performed well, we observed occasional under-specification of rules, especially in more dynamic communication scenarios. 
These cases highlight areas where postprocessing may be beneficial to ensure complete policy coverage.
\noindent\textbf{Role Refinement:}~\label{par:llama_role_refinement}
Llama achieved high precision ($0.975$) and strong recall ($0.758$), resulting in an F1 of $0.808$. 
It reliably removed unnecessary permissions, but sometimes missed edge-case access rules in ambiguous or infrequent access patterns.
These results indicate effective RBAC refinement capabilities with room for improvement in recall.
\noindent\textbf{NetworkPolicy Refinement:}~\label{par:llama_networkPolicy_refinement}
Llama exhibited mixed performance, with a relatively high precision of $0.889$, but a low recall of $0.385$, leading to an F1 of $0.504$.
This low recall was due to the model often missing service connections, a result of the complexity of reconstructing network interactions from dense log data. 
Nonetheless, even partial refinements contribute to reducing the overall attack surface.
Given these limitations, we suggest complementing NetworkPolicy Refinement with NetworkPolicy Creation to uncover additional traffic flows and improve overall security assurance.
\noindent\textbf{Deployment Refinement:}~\label{par:llama_deployment_refinement}
In this task, Llama exhibited moderate performance, achieving a precision of $0.689$, a recall of $0.804$, and an F1 of $0.728$. 
The high recall shows that the model successfully captures most necessary refinements, while the moderate precision indicates it also includes some irrelevant or unnecessary changes.
Although Llama's precision is moderate, its high recall indicates that it effectively leverages APLs to propose context‑aware, container‑level security improvements.

Processing time results further confirmed SLMs' practicality, with all Llama-3.1-8B prompt chains completing within usable limits (up to 15 minutes), supporting their integration as offline adaptive hardening workflows (complete results in Table~\ref{tab:appendix-processing-time}).

In this section (\ref{subsec:evaluating_the_best_local_SLM_Llama}), we evaluated how well an SLM performs when applied to various K8s resource creation and refinement tasks. 
As expected, the SLM performs relatively well, though not as well as the proprietary LLM (GPT-4o, Sec.~\ref{subsec:evaluating_the_best_proprietary_LLM_GPT4o}). 
The following sections,~\ref{subsec:iterative_security_refinement} and~\ref{subsec:extending_from_local_slm_to_local_llm_qwen2.5_14b}, present two strategies that we developed and evaluated to improve the performance of SLMs and support use cases where organizations must rely on a local open-source model.

\subsection{Iterative Security Refinement}\label{subsec:iterative_security_refinement}

\begin{figure}[h]
    \centering
    \includegraphics[width=0.85\linewidth, trim=11 25 11 7, clip]{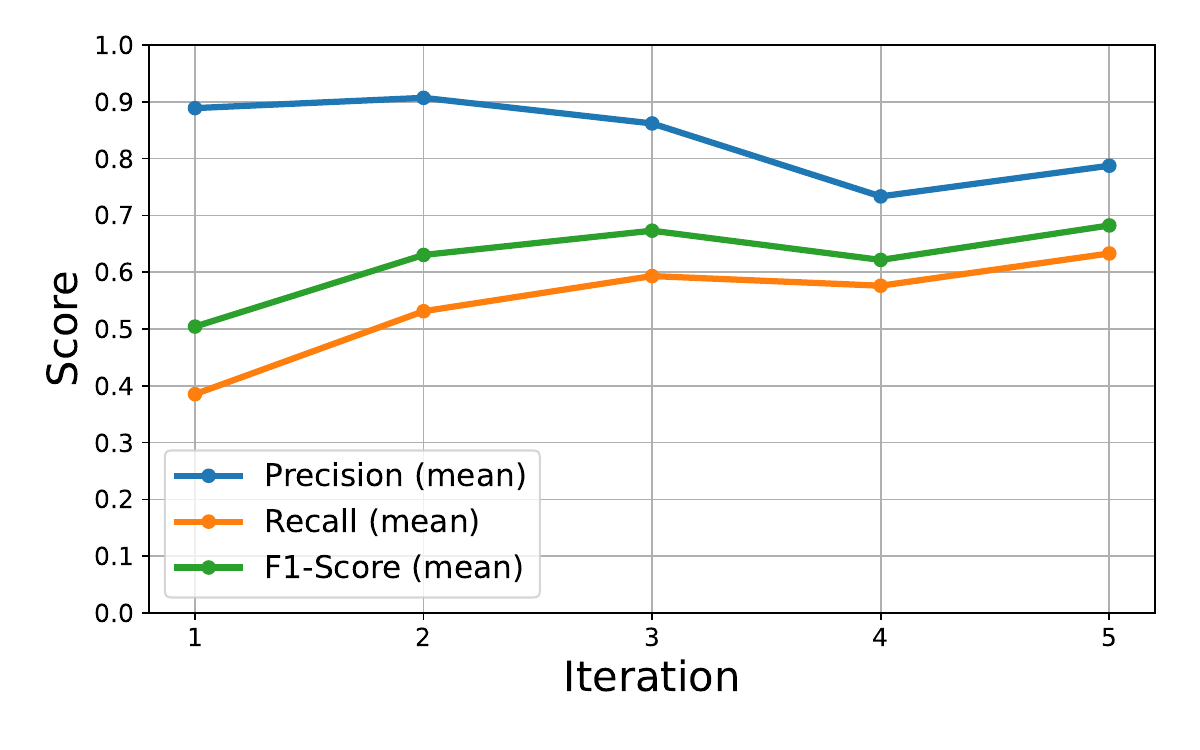}
    \caption{Average precision, recall, and F1-score across five NetworkPolicy Refinement iterations.}
    \label{fig:avg_metrics_refinement_iterations}
\end{figure}

As discussed in Sec.~\ref{par:llama_networkPolicy_refinement}, Llama initially showed low recall ($0.385$) in the NetworkPolicy Refinement task, resulting in a moderate F1 of $0.504$. 
To address this, we developed an iterative refinement loop that recursively feeds the output of each prompt chain iteration (a refined manifest) back into the model as input for the next.
This enables the model to incrementally improve its recommendations by building on previously refined configurations.
In our experiments, shown in Fig.~\ref{fig:avg_metrics_refinement_iterations}, NetworkPolicies refined by the SLM became increasingly aligned with observed traffic restrictions over the first three iterations:
Recall and F1 improved consistently, from $0.385$ to $0.593$, and from $0.504$ to $0.673$, respectively, and 
precision remained consistently high, only slightly decreasing from $0.889$ to $0.862$. 
However, performance declined across all metrics in the fourth iteration, indicating that the model over-pruned some legitimate rules.
To verify whether the decline was persistent, we ran a fifth pass, which lifted precision to $0.787$, recall to $0.633$, and F1 to $0.682$.
This indicates that the fourth iteration drop was primarily a temporary effect 
and not a progressive performance drift. 
The net gain in F1 beyond the third iteration and the alternating drop-recovery pattern reveal an oscillation with diminishing returns, suggesting that continued refinement beyond this point may be unnecessary.
In general, iterative refinement improves manifest security through five iterations (cumulative increase of $0.178$ in F1). 
The only potential drawback is oscillations in later rounds, which a simple early stopping criterion can readily mitigate.

\subsection{Extending from Local SLM to Local LLM}\label{subsec:extending_from_local_slm_to_local_llm_qwen2.5_14b}

To explore whether increasing model size improves performance, we evaluated a larger open-source LLM, Qwen2.5-14B~\cite{qwen2}.
Despite Llama-3.1-8B being the top-performing SLM in earlier experiments (Sec.~\ref{subsec:selecting_best_local_os_slms}), its next‐size variant has 70 billion parameters~\cite{meta-llama-llama3.1-70B-instruct}, making it impractical for many local deployments and common users. 
Consequently, we selected Qwen2.5-14B~\cite{qwen2} as the larger representative local LLM to compare directly with its smaller variant, Qwen2.5-7B, 
for a fair evaluation of model size.
We focused on two tasks that provide a strong benchmark for scaling effects: NetworkPolicy Creation (previously used as baseline in Sec.~\ref{subsec:selecting_best_local_os_slms}) and NetworkPolicy Refinement, which targets the same resource but is more complex and proved challenging for Llama-3.1-8B (required reasoning across multiple chain outputs in Sec.~\ref{subsec:iterative_security_refinement}).

In NetworkPolicy Creation, Qwen2.5-14B achieved precision of $0.781\pm0.172$ and recall of $0.774\pm0.171$, yielding F1 of $0.774\pm0.166$. 
Although this performance was slightly below Llama-3.1-8B, which reached F1 of $0.793$, it still marked an improvement over its smaller counterpart, Qwen2.5-7B, in all metrics (Table~\ref{tab:experimenting-slms-netpol-creation}). 
These results suggest that model alignment may be the cause of the performance gap observed with Llama. 
More notably, in NetworkPolicy Refinement, Qwen2.5-14B reached precision of $0.724\pm0.316$, recall of $0.645\pm0.352$, and F1 of $0.667\pm0.340$. 
This significantly outperformed Llama-3.1-8B, which only achieved an F1 of $0.504$ with a recall of $0.385$ (Table~\ref{tab:llama-performance}).
These findings indicate that scaling to a larger local LLM can yield measurable improvements in certain hardening tasks, particularly refinement.
Thus, Qwen2.5-14B offers a promising alternative when additional model capacity is available.

\subsection{Empirical Comparison to Existing Methods}\label{subsec:Empirical_Comparison_to_Existing_Methods}
We evaluated \name with two methods that perform identical tasks, ensuring a valid empirical comparison:
audit2rbac~\cite{liggitt_audit2rbac_2023} and KUBETEUS~\cite{kimkubeteus} (reviewed in Sec.~\ref{subsec:secure_k8s_config_generation}). 
audit2rbac generates Roles and RoleBindings derived from K8s audit logs and API requests for each user. 
KUBETEUS employs intent-driven LLM prompts on static manifests and live API state to generate NetworkPolicies, and was also evaluated on \textit{Online Boutique}~\cite{google_microservices_demo}, one of our testbed applications. 
We evaluated the performance of \name against each method with the same workloads, quantifying its relative effectiveness under identical conditions.

We compared \name (using GPT-4o) with audit2rbac on Role Creation (Table~\ref{tab:audit2rbac-empricial-comparison}).
While user-centric audit2rbac achieved perfect recall, it overgenerated rules, thus producing many FPs and resulting in low precision ($0.208$) and F1 ($0.343$). 
In contrast, the pod-centric approach of \name achieved perfect scores.

We evaluated \name against KUBETEUS on NetworkPolicy Creation under the same constrained compute conditions, using the Deepseek-Coder-1.3B SLM~\cite{guo2024deepseekcoderlargelanguagemodel} (Table~\ref{tab:kubeteus-empricial-comparison}). 
Both methods showed modest performance due to the model's limited capacity, but \name outperformed KUBETEUS, achieving higher precision ($0.444$ vs. $0.360$), recall ($0.265$ vs. $0.238$), and F1 ($0.332$ vs. $0.275$).

Through this empirical comparison, we demonstrate how \name, grounded in Pod-aggregated audit and network data, yields significantly more secure manifests across both Roles and NetworkPolicies, compared to other relevant, task-equivalent tools.

\subsection{Sensitivity Analysis}\label{subsec:sensitivity_analysis}

\subsubsection{Prompt Chain Order}\label{subsubsec:sens_prompt_chain_order}

To assess how prompt ordering affects performance in chain-of-prompts pipelines, we conducted a sensitivity analysis on Deployment Refinement.
This task was chosen because it relies on both AALs and APLs, making it one of \name's longest and most complex prompt chains.
The analysis evaluated how varying the order of functional prompts affects model output.
We empirically evaluated three prompt chain orders. 

As shown in Table~\ref{tab:chain-order-sensitivity-analysis}, the order in which prompts are chained has a noticeable effect on performance.
$Manifest \rightarrow Logs$ refers to a reverse chaining strategy, where the model is first given the manifest, followed by the aggregated logs.
This contrasts with our original method, $Logs \rightarrow Manifest$, which begins by analyzing behavioral logs before the manifest. 
The reversed configuration achieved an F1 of $0.909$, trailing slightly behind the original method's $0.929$.
Unlike our approach, which first analyzed all data sources and then determined their relationship to the manifest, $Analyze\&Match$ analyzed each data source individually and matched it immediately with its corresponding K8s resource.
The $Analyze\&Match$ configuration yielded the lowest F1 ($0.848$), mainly due to reduced recall ($0.815$).
These findings indicate that prompt sequencing is a key factor in \name's effectiveness. 
Designing prompt chains to reflect real-world reasoning, by interpreting logs before updating the manifests, improves model alignment and output accuracy.

\begin{table}[h]
\centering
\caption{Performance of different prompt chain orders on Deployment Refinement.}
\resizebox{0.9\columnwidth}{!}{
\begin{tabular}{|l|l|l|l|}
\hline
\textbf{Chain Order} & \textbf{Precision} & \textbf{Recall} & \textbf{F1-Score} \\ \hline
$Logs \rightarrow Manifest$
& $0.975 \pm 0.053$ & $0.894 \pm 0.140$ & $0.929 \pm 0.100$ \\ \hline
$Manifest \rightarrow Logs$
& $0.973 \pm 0.063$ & $0.864 \pm 0.157$ & $0.909 \pm 0.109$ \\ \hline
$Analyze \& Match$
& $0.905 \pm 0.145$ & $0.815 \pm 0.251$ & $0.848 \pm 0.217$ \\ \hline
\end{tabular}
}
\label{tab:chain-order-sensitivity-analysis}
\end{table}

\subsubsection{Sensitivity To Data Collection Duration}\label{subsubsec:data_logs_sens}

To evaluate how the duration of data collection affects the completeness of aggregated logs, we replicated the first step of our method (``Data Acquisition and Preprocessing'', Sec.~\ref{subsec:data_acquisition_and_preprocessing}) to produce AALs, ANLs, and APLs of varying durations for comparison. 
Each demo application was run with its built-in load generator~\cite{microservicesdemo_loadgenerator,aksstoredemo_virtualcustomer,aksstoredemo_virtualworker} for seven consecutive days to collect logs accordingly (Fig.~\ref{fig:proposed_method}(a)).
The resulting trace was then divided into 100 cumulative segments: segment~1 contains only the earliest events, segment~2 contains all events from segment~1 plus the next segment, and so on, until segment~100 covers the full week.
Each cumulative segment was processed to generate AALs, ANLs, and APLs using K-V aggregation (Algorithm ~\ref{alg:log_aggregation}).
We then measured the similarity between consecutive segment pairs~($\langle\text{segment}_{i},\text{segment}_{i+1}\rangle$) for each type of aggregated log using Cosine, Overlap, and Dice similarity metrics (Sec.~\ref{subsubsec:ablation_effect _of _data_explanations}).
This time-lag analysis reveals how quickly each aggregated log type converges as the duration of data collection increases.
\begin{figure}[h]
  \centering
  \includegraphics[width=\linewidth, trim=5 5 8 8, clip]{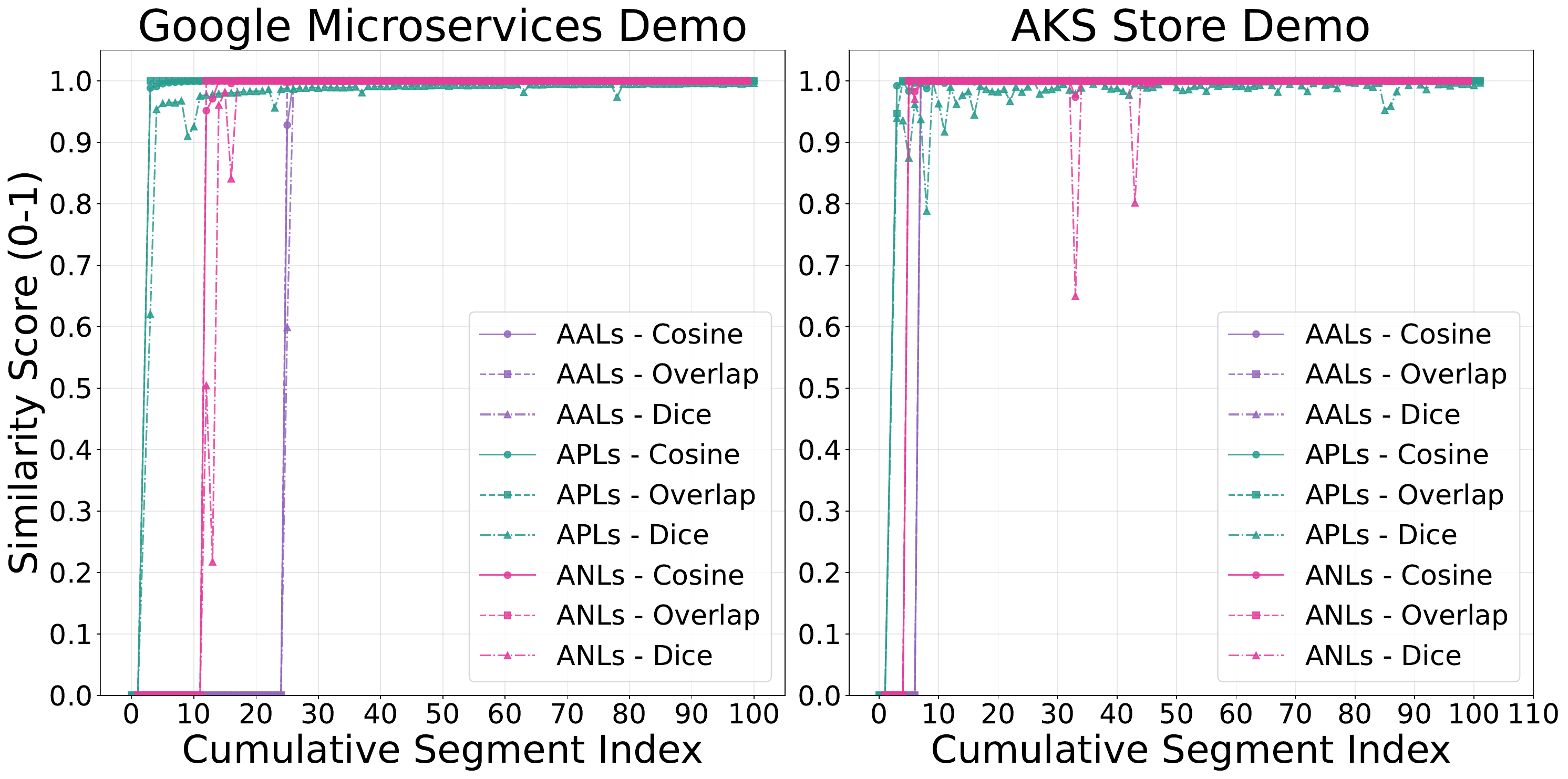}
  \caption{Similarity between successive cumulative aggregated log segments across log types and applications.}
  \label{fig:agg_similarity_consecutive_segments}
\end{figure}
Each point in Fig.~\ref{fig:agg_similarity_consecutive_segments} represents the similarity between two consecutive segments: A value of 1 indicates identical content, while lower values reflect changes. 
The figure includes two graphs, one for each application's results (Google's ``Microservices Demo''~\cite{google_microservices_demo} and``AKS Store Demo''~\cite{azure_store_demo}), allowing comparison of how quickly each application's logs converge. 
We observed that all three log types rapidly approached convergence, reaching a similarity score of almost $1.0$ early in the trace, with each type converging at a different point in time.
The similarity of APLs and ANLs stabilized in about 4 hours (reaching a similarity of approximately $0.98$) while AALs required a longer horizon of about $1.75$ days to reach the same score.
Beyond these points, longer runs did not contribute meaningful changes, indicating that aggregation captures and preserves core behavioral patterns early. 
Thus, once convergence is reached, the duration of data collection no longer impacts the aggregated output, which remains stable across both applications under realistic workloads.

Notably, aggregation reduced token volume by around $5.6$ million tokens ($99.96\%$) across log types on our evaluation datasets, composed of log data collected after the convergence point, when additional data no longer introduced new information.
This substantial compression shows that collecting logs beyond convergence merely inflates the dataset size by millions of tokens without benefit. 
This efficiency keeps input data within LLM token limits, reduces cost, and improves accuracy.
It is important to note that operators should generate representative traffic to ensure adequate coverage before convergence.
Moreover, the results show that K-V aggregation helps mitigate sensitivity to data sufficiency, ensuring reliable results even when exact data completeness varies.

\subsection{Ablation Study}\label{sec:ablation_Study}

\subsubsection{Effect of Data Explanations}\label{subsubsec:ablation_effect _of _data_explanations}

To assess the role of explanatory context, we reversed \name's default settings by removing log type descriptions where they were originally used and adding them where they were absent.
As shown in Table~\ref{tab:ablation_study_data_explanations}, Deployment Refinement (Sec.~\ref{bullet:Deployment_refinement}) performance decreased without additional explanations about the data, with F1 of $0.881$ compared to $0.929$ in the original configuration (Table~\ref{tab:gpt4o-performance}). 
This reduction suggests that the task benefits from added context, as explanations help the model integrate multiple log sources (AALs and APLs) in more complex tasks for more accurate refinements.
Conversely, adding explanations reduced performance in simpler tasks like Role Creation (F1 of $1.00$ vs. $0.735$).
These declines suggest that when input requirements are minimal, added context can introduce unnecessary noise that disrupts the model’s reasoning.
Overall, the results suggest that data explanations benefit complex tasks involving multi-source integration but may hinder performance in simpler tasks, where concise inputs better support model reasoning.

\begin{table}[h]
\centering
\caption{GPT-4o performance across prompt chains with and without additional data explanations.}
\resizebox{\columnwidth}{!}{
\begin{tabular}{|l|l|l|l|l|}
\hline
\textbf{Task} & \textbf{Explanations} & \textbf{Precision} & \textbf{Recall} & \textbf{F1-Score} \\ \hline
Role Creation & With 
& $0.776 \pm 0.327$ & $0.844 \pm 0.223$ & $0.735 \pm 0.250$ \\ \hline
NetworkPolicy Creation & With 
& $0.938 \pm 0.091$ & $0.871 \pm 0.156$ & $0.897 \pm 0.118$ \\ \hline
Role Refinement & With 
& $1.00 \pm 0.0$ & $0.860 \pm 0.155$ & $0.917 \pm 0.100$ \\ \hline
NetworkPolicy Refinement & With 
& $0.952 \pm 0.061$ & $0.889 \pm 0.138$ & $0.916 \pm 0.101$ \\ \hline
Deployment Refinement & Without 
& $0.949 \pm 0.107$ & $0.826 \pm 0.119$ & $0.881 \pm 0.105$ \\ \hline
\end{tabular}
}
\label{tab:ablation_study_data_explanations}
\end{table}

\subsubsection{Effect of Removing the Matching Step from Refinement Prompt Chains}\label{subsubsec:ablation_removing_matching_step_from_chain}

We tested how removing the \textit{matching} step from refinement prompt chains impacts performance, expecting degradation.
Table~\ref{tab:ablation_study_removing_match_sub_task} shows that removing it reduced performance for all tasks:
In Role Refinement, recall dropped from $0.914$ to $0.721$, lowering F1 to $0.835$; 
NetworkPolicy Refinement saw a larger decline, with F1 dropping from $0.961$ to $0.652$;
Deployment Refinement was less affected, but its performance still decreased (F1 of $0.929$ vs. $0.854$).
These results highlight the critical role of the \textit{matching} step in aligning log-derived insights with K8s manifests for refinement chains, facilitating more accurate context-aware recommendations. 

\begin{table}[h]
\centering
\caption{
GPT-4o performance across refinement tasks after removing the matching step from prompt chains.}
\resizebox{\columnwidth}{!}{
\begin{tabular}{|l|l|l|l|}
\hline
\textbf{Task (No \textit{Matching})} & \textbf{Precision} & \textbf{Recall} & \textbf{F1-Score} \\ \hline
Role Refinement 
& $1.00 \pm 0.0$ & $0.721 \pm 0.075$ & $0.835 \pm 0.048$ \\ \hline
NetworkPolicy Refinement
& $0.791 \pm 0.298$ & $0.562 \pm 0.223$ & $0.652 \pm 0.247$ \\ \hline
Deployment Refinement
& $0.924 \pm 0.109$ & $0.810 \pm 0.163$ & $0.854 \pm 0.124$ \\ \hline
\end{tabular}
}
\label{tab:ablation_study_removing_match_sub_task}
\end{table}

\subsubsection{Effect of Omitting Data Sources in Deployment Refinement}\label{subsubsec:ablation_omitting_data_sources_from_deployment_refinement}

To evaluate the importance of each data source in Deployment Refinement, we removed AALs and APLs separately. 
Compared to the original configuration, which achieved an F1 of $0.929\pm0.100$ (Sec.~\ref{subsec:evaluating_the_best_proprietary_LLM_GPT4o}), removing AALs had the strongest effect, lowering F1 to $0.829\pm 0.137$, having a precision of $0.928\pm0.088$, and a reduced recall of $0.765\pm 0.177$.
Omitting APLs led to a smaller drop—F1 of $0.876\pm0.149$, as \name maintained a precision of $0.938\pm0.116$ and a recall of $0.831\pm0.177$.
These results underscore that both AALs and APLs contribute uniquely valuable context, and their combined use is essential for accurate and complete manifest hardening, highlighting the value of multi-source integration.

\subsection{Error Analysis}\label{subsec:error_analysis}

To better understand \name's performance and where it struggles, we conducted detailed error analysis for all hardening tasks.

GPT-4o’s errors were rare, interpretable, and often benign, such as configuring source ports instead of destination ports for high-traffic NetworkPolicies, which can be corrected with postprocessing.
However, since overall performance is already high, adding more steps to prompt chains would lengthen the pipeline and risk introducing noise without meaningful gains.
\textit{Role Refinement} performed well in general. 
Its main error, retaining rare infrastructure‑facing permissions, reflects a cautious stance to preserve functionality, hence acceptable.
Similarly, \textit{NetworkPolicy Refinement} had minor recall drops due to missed connections in high‑traffic services, underscoring challenges of reasoning over large‑volume logs.
This can be mitigated through postprocessing as well (e.g., rule-based or human-in-the-loop).
\textit{Deployment Refinement} also performed well overall, while its main issue was pretrained LLMs' tendency to add beneficial best‑practice hardenings, although unrequested~\cite{zhou2024comprehensive,yang2024harnessing,huang2025survey}.

Llama-3.1-8B, while smaller, demonstrated high precision across tasks. 
However, like GPT‑4o, it sometimes under-specified ingress or egress rules in traffic-heavy NetworkPolicy refinements, but also missed rare permissions in Role refinements, likely due to low frequency and ambiguous usage, leading to slightly lower recall. 
In \textit{Role Creation}, it occasionally added unobserved actions, for example, permissions for persistent storage resources (e.g., PersistentVolumes~\cite{kubernetes_persistent_volumes}), even when no storage-related activity was detected. 
Although potentially security-enhancing, these were out of \name's scope; therefore, they reduced precision.
In general, errors reflected trade-offs between log adherence and operational security, rather than systemic flaws.
\name's iterative refinement and human oversight help resolve such edge cases.

\section{Limitations}\label{sec:Limitations}

While \name provides an effective approach to K8s security hardening, it has several limitations.
First, although \name supports both Resource Creation and Refinement, it currently targets only Roles, NetworkPolicies, and Deployments. 
Future work aims to broaden this scope.
Second, although our modular prompt chaining has proved effective across tasks and resources, crafting prompts requires expertise, particularly when adapting to new resource types or domains.
Third, although \name generates ready-to-use hardened manifests, it delegates enforcement to the user.
This design prioritizes safety and transparency, but may hinder adoption when automated remediation is required.
Regarding strategies for improving SLM performance, although our results are promising, future work could explore adaptive stopping for iterative refinement in depth for further improvement.
Notably, capturing sufficient log data requires representative traffic, often a limitation for log-driven approaches, similar to dynamic analysis~\cite{li2021automatic,clausen2019traffic,xiang2019towards}.
This can be mitigated through either load generators that exercise relevant endpoints (e.g., Locust~\cite{locust,microservicesdemo_loadgenerator}, Rust-based~\cite{rust,aksstoredemo_virtualcustomer,aksstoredemo_virtualworker}), or by recording real user traffic.
Since \name can be applied offline and in any standard deployment environment, such activity can be produced without disrupting operational workflows.
Moreover, our K-V aggregation method ensured sufficient coverage in our testbed environment (Sec.~\ref{subsubsec:data_logs_sens}).

\section{Conclusion}\label{sec:conclusion}

In this research, we introduced \emph{\name}, an LLM-guided log-driven framework that leverages prompt chaining to transform dynamic observability data into actionable least-privileged K8s manifests. 
By supporting both Resource Creation and Refinement, \name serves as a practical user-assistive solution for scalable hardening in modern threat-prone environments.
By integrating audit, network, and provenance logs into structured prompt chains, \name generates precise, context-aware, least-privilege manifests that reduce the attack surface.
Comprehensive evaluations demonstrated \name's effectiveness. 
The best results were achieved by the proprietary GPT-4o model, while among open-source SLMs, Llama-3.1-8B demonstrated solid performance, making it suitable for cost-sensitive or privacy-focused settings.

While \name performs well for core K8s resources, future work will expand its coverage to include additional resource types, such as ConfigMaps~\cite{kubernetes_configmap} and PersistentVolumes~\cite{kubernetes_persistent_volumes}.
Extending resource support will provide a more comprehensive hardening solution that spans the entire K8s cluster lifecycle.
In addition, future work will evaluate larger open-source LLMs (e.g., Llama-3.1-70B/405B~\cite{grattafiori2024llama}, DeepSeek-V3-671B~\cite{deepseekai2024deepseekv3technicalreport}) to assess whether greater capacity and reasoning ability improve performance, especially for complex refinement tasks, while preserving the benefits of local privacy-aware inference.
Lastly, future research will integrate application-level context, such as system architecture and component roles, to facilitate more targeted security recommendations.

\bibliographystyle{ACM-Reference-Format}
\bibliography{references}

\appendix

\section{Example Deployment Manifest}\label{app:manifest_example}

Listing~\ref{lst:nginx_manifest} provides an example of a simple K8s Deployment manifest, named \texttt{nginx-deployment}, which runs two replicas of the official \texttt{nginx:1.14.2} container image. 
Each container exposes port 80. 
This example demonstrates the typical structure of a Deployment object, including metadata, selectors, and Pod specifications.

\begin{minipage}{0.7\linewidth}
\begin{tcolorbox}[
    boxrule=0.3pt,
    arc=2pt,
    left=2mm, right=2mm, top=1mm, bottom=1mm,
    nobeforeafter
]
\footnotesize
\begin{lstlisting}[label={lst:nginx_manifest}]
apiVersion: apps/v1
kind: Deployment
metadata:
  name: nginx-deployment
spec:
  replicas: 2
  selector:
    matchLabels:
      app: nginx
  template:
    metadata:
      labels:
        app: nginx
    spec:
      containers:
      - name: nginx
        image: nginx:1.14.2
        ports:
        - containerPort: 80
\end{lstlisting}
\end{tcolorbox}
\end{minipage}

\section{K-V Log Aggregation Algorithm}\label{app:log_aggregation_algorithm}

Algorithm~\ref{alg:log_aggregation} outlines the process of log aggregation used in \name to reduce the input size for downstream LLM-based analysis while preserving semantic relevance.
This recursive algorithm takes a nested JSON log object as input and aggregates values for each key into sets, effectively flattening the structure and reducing redundancy. 
The result is a mapping from each key to a set of unique observed values across the input structure.
The algorithm defines a recursive function \texttt{Aggregate}, which traverses the JSON object:
If a key maps to a primitive value (e.g., string, number, or boolean), the value is added to the set associated with that key. 
If the value is a nested dictionary or list, the function recurses into the structure.
Lists are treated as unordered sequences of items to recursively aggregate.
The final output is a HashMap where each key from the original (possibly deeply nested) log is mapped into a set of unique values. 
This representation reduces verbosity and repetition, making it suitable for token-efficient LLM input.
This aggregation process is applied to logs at the preprocessing stage (Sec.~\ref{subsec:data_acquisition_and_preprocessing}) to distill log data while retaining its semantic footprint. 
It supports coherent downstream interpretation by LLMs under token and context constraints.

\begin{algorithm}[h]
\caption{KeyValueAggregationFromNestedJSON}\label{alg:log_aggregation}
\SetKwFunction{FRec}{Aggregate}
\KwIn{A JSON object $J$}
\KwOut{A dictionary $A$ mapping each key to a \texttt{set} of values}
$A \leftarrow$ empty map of sets\;
\FRec{$J$, $A$}\;
\Return $A$\;
\vspace{0.2cm}
\SetKwProg{Fn}{Function}{:}{}
\Fn{\FRec{$x$, $A$}}{
  \If{$x$ is a dictionary}{
    \ForEach{key, value in $x$}{
      \If{$value$ is primitive}{
        $A[key] \leftarrow A[key] \cup \{value\}$\;
      }
      \Else{
        \FRec{$value$, $A$}\;
      }
    }
  }
  \If{$x$ is a list}{
    \ForEach{$item$ in $x$}{
      \FRec{$item$, $A$}\;
    }
  }
}
\end{algorithm}

\section{Aggregated Logs Illustrations and Examples}

\begin{figure}[h]
    \centering
    \includegraphics[width=\linewidth, trim=20 20 20 10, clip]{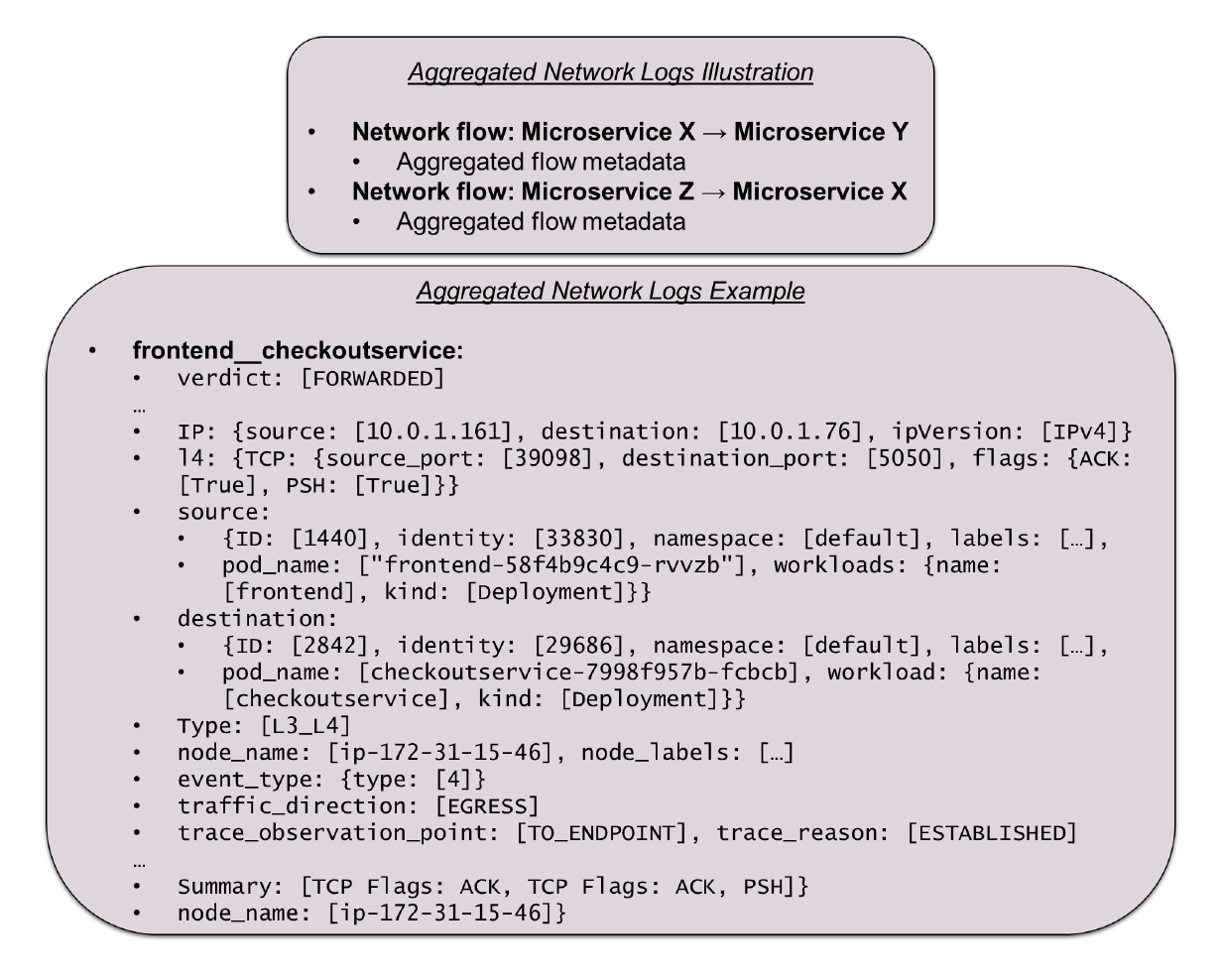}
    \caption{Illustration and example of Aggregated Network Logs (ANLs).}
    \label{fig:ANL_example}
\end{figure}

\paragraph{ANLs}
Fig.~\ref{fig:ANL_example} shows an illustration and an example of Aggregated Network Logs (ANLs, Sec.~\ref{subsubsec:Aggregated_Network_Logs}). 
The illustration outlines the network aggregation process, where Pod-level communication flows are captured using Hubble~\cite{hubble_observability,hubble_github} and grouped by communicating service pairs. 
The example output demonstrates ANLs' structure used in \name, in which network events (e.g., ingress and egress connections, ports, and protocols) are stored as deduplicated key-value mappings. 
This representation enables precise mapping between observed network behavior and declared NetworkPolicy rules, forming the basis for policy hardening while retaining only essential service-relevant communication data.

\begin{figure}[h]
    \centering
    \includegraphics[width=\linewidth, trim=20 20 20 20, clip]{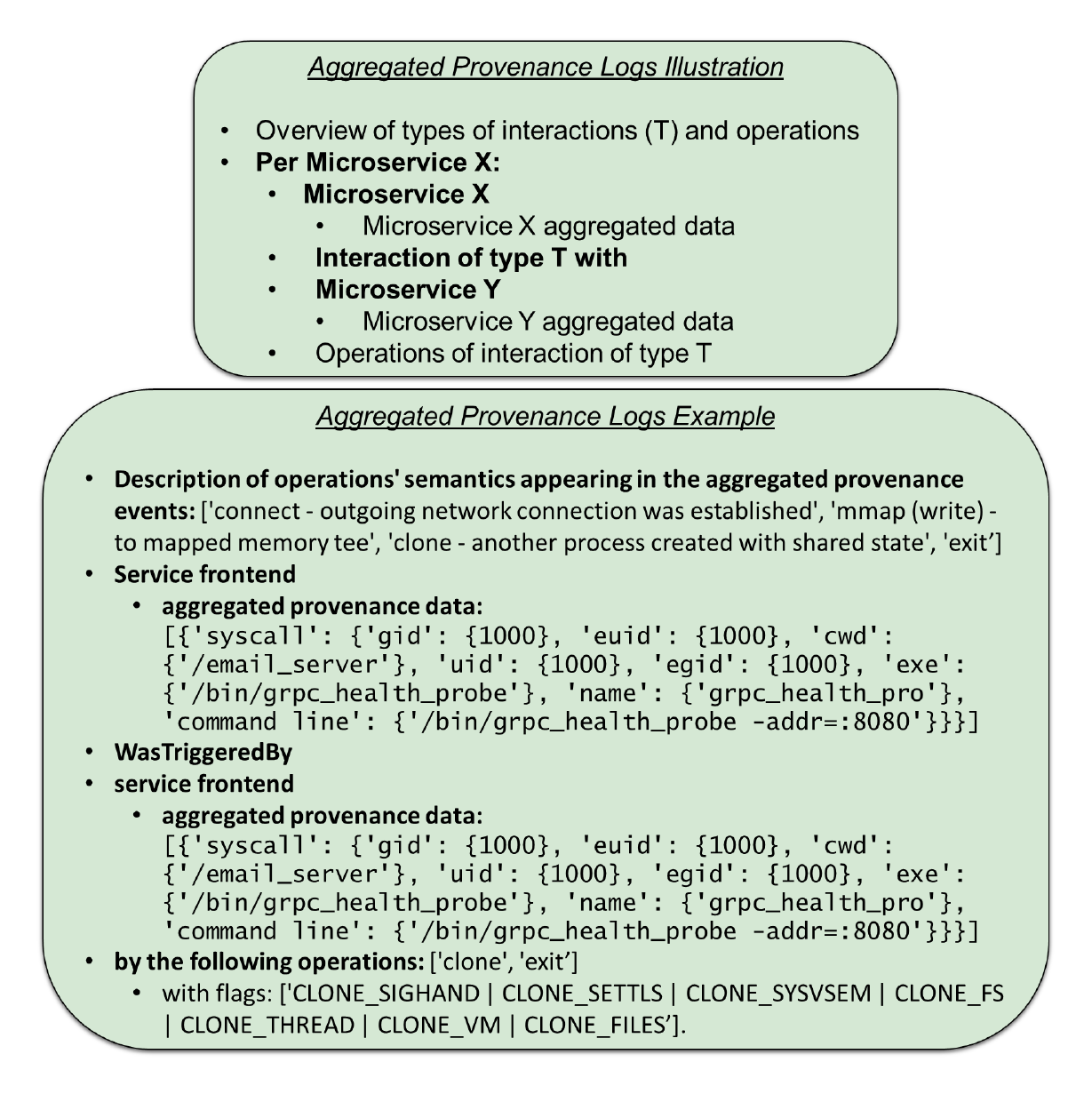}
    \caption{Illustration and example of Aggregated Provenance Logs (APLs).}
    \label{fig:APL_example}
\end{figure}

\paragraph{APLs}
Fig.~\ref{fig:APL_example} presents an illustration and an example of Aggregated Provenance Logs (APLs, Sec.~\ref{subsubsec:Aggregated_Provenance_Logs}). 
The illustration depicts the provenance aggregation process, where low-level system events, such as file operations, process creations, and inter-process communications, are collected by SPADE~\cite{gehani2012spade,SPADE} and mapped to application-level microservices. 
The example output shows APLs' representation used in \name, in which provenance graph vertices and edges relevant to a specific service are condensed into key–value records. 
This structured format preserves essential causal relationships while reducing log volume, enabling efficient manifest analysis and hardening (e.g., for Deployments).

\section{Prompting Techniques Examples}

\begin{figure}[h]
    \centering
    \includegraphics[width=\linewidth, trim=15 20 20 20, clip]{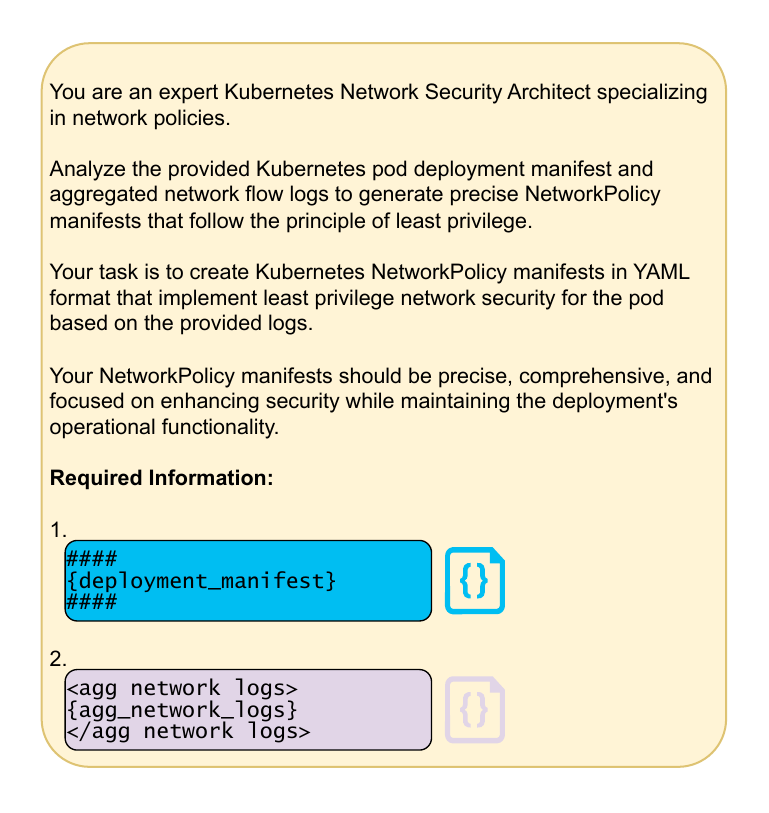}
    \caption{Zero-shot prompt example of the NetworkPolicy Creation task.}
    \label{fig:zero_shot_example}
\end{figure}

\paragraph{Zero-Shot} 
Fig.~\ref{fig:zero_shot_example} presents an example of a zero-shot prompt used for the \textit{NetworkPolicy Creation} task. 
In this approach, the language model receives a direct instruction, without intermediate examples or reasoning-structured context, to generate K8s NetworkPolicy manifests based solely on the input Deployment manifest and aggregated network logs (ANLs).
The prompt includes the relevant K8s Pod specification and a summary of the Pod’s observed communication patterns. 
The LLM is expected to analyze this information and generate precise NetworkPolicy rules that follow the principle of least privilege, ensuring secure inter-Pod communication while maintaining application functionality.
Zero-shot prompting serves as a baseline for evaluating LLMs' capacity to understand task intent and generate correct outputs without auxiliary guidance or step-by-step reasoning. 
While effective in simpler scenarios, it may lack the robustness and specificity provided by more advanced prompting techniques, such as chain-of-thought or prompt-chaining strategies.

\begin{figure}[h]
    \centering
    \includegraphics[width=1.1\linewidth]{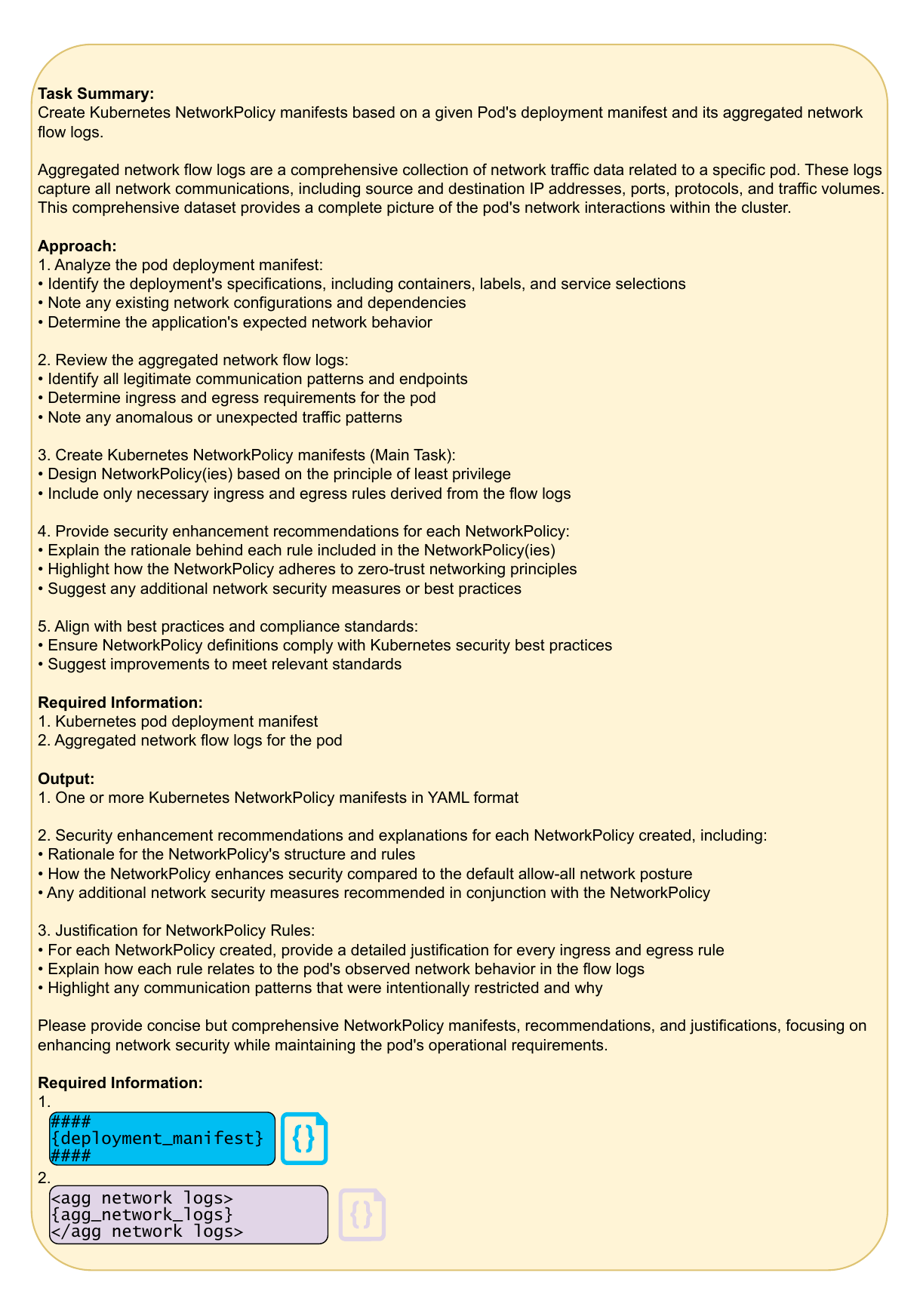}
    \caption{Chain-of-thought prompt example of the NetworkPolicy Creation task.}
    \label{fig:cot_example}
\end{figure}

\paragraph{Chain-of-Thought} Fig.~\ref{fig:cot_example} illustrates a chain-of-thought (CoT) prompt used for the \textit{NetworkPolicy Creation} task. 
Unlike zero-shot prompting, the CoT prompt explicitly guides the LLM through a step-by-step reasoning process to support the generation of precise, security-focused Kubernetes NetworkPolicy manifests.
The prompt outlines a structured procedure: first, analyzing the Pod Deployment manifest; second, reviewing aggregated network logs (ANLs) to identify legitimate communication patterns; third, designing least-privilege NetworkPolicy rules; and finally, explaining and justifying each rule. 
This decomposition helps the model internalize the logical flow required to align security configurations with real-world application behavior.
The CoT strategy improves interpretability and reduces the likelihood of errors or omissions by breaking the task into smaller, manageable reasoning steps. 
This makes it particularly useful for complex scenarios where understanding context and explaining policy decisions is essential to ensuring correctness and trust in the generated outputs.

\section{Open-Source Local SLM Processing Time Results}\label{app:processing-time-statistics}

We monitored the processing time of each prompt chain powered by the best-performing local SLM, Llama-3.1-8B (Sec.~\ref{subsec:evaluating_the_best_local_SLM_Llama}).
The results show that each chain completes within practical bounds: 
Role Creation finishes in a median of 4.69 minutes (min),
NetworkPolicy Creation in a median of 8.46 min,
Role Refinement in a median of 10.78 min,
NetworkPolicy Refinement in a median of 12.91 min,
and Deployment Refinement in a median of 11.31 min.
In addition, even in the case of occasional long runs, all prompt chains remained under a 5-hour ceiling, and most have completed in under 15 minutes, affirming their suitability for adaptive security-hardening workflows.

Table \ref{tab:appendix-processing-time} lists the complete aggregated processing time statistics (all values in minutes) for each prompt chain. 
We report the mean, minimum, maximum, median (P50), 75th, 90th, 95th, and 99th percentiles, as well as the interquartile range (IQR).
We define these statistics as follows:

\begin{description}[noitemsep,leftmargin=1em]
  \item[\textbf{Mean:}] Arithmetic average of all observed processing times; conveys the expected overall latency.
  \item[\textbf{Min:}] Shortest observed processing time; indicates the best‐case performance.
  \item[\textbf{Max:}] Longest observed processing time; indicates the worst‐case or tail latency.
  \item[\textbf{P50 (Median):}] 50th percentile; the processing time at which half of the runs complete faster and half slower, robust to outliers.
  \item[\textbf{P75:}] 75th percentile; 75\% of runs complete within this time, highlighting the upper‐middle range.
  \item[\textbf{P90:}] 90th percentile; useful to gauge how often runs approach the longer tail.
  \item[\textbf{P95:}] 95th percentile; captures near‐worst‐case latency without being dominated by extreme outliers.
  \item[\textbf{P99:}] 99th percentile; illustrates the very-long‐tail behavior critical for worst‐case planning.
  \item[\textbf{IQR (Interquartile Range):}] Difference between P75 and P25; measures variability around the median and the spread of the middle 50\% of runs.
\end{description}

\begin{table}[h]
  \centering
  \caption{Aggregated Llama-3.1-8B processing times statistics for each prompt‐chain, in minutes.}
  \label{tab:appendix-processing-time}
  \resizebox{\columnwidth}{!}{%
    \begin{tabular}{|l|r|r|r|r|r|r|r|r|r|}
      \hline
      \textbf{Prompt Chain}  & \textbf{Mean}  & \textbf{Min}   & \textbf{Max}    & \textbf{P50}   & \textbf{P75}   & \textbf{P90}    & \textbf{P95}    & \textbf{P99}    & \textbf{IQR}   \\
      \hline
      Role Creation                    
      &  4.69 &  4.68 &   4.70 &  4.69 &  4.69 &   4.70 &   4.70 &   4.70 &  0.00 \\
      \hline
      NetworkPolicy Creation           
      &  9.91 &  5.44 &  20.20 &  8.46 & 9.14 &  19.89 &  20.20 &  20.20 &  3.14 \\
      \hline
      Role Refinement                  
      & 27.88 &  7.32 & 193.04 & 10.78 & 18.15 &  89.49 & 124.84 & 160.48 & 13.46 \\
      \hline
      NetworkPolicy Refinement         
      & 40.07 &  6.70 & 290.22 & 12.91 & 41.99 & 175.41 & 204.15 & 290.22 & 32.72 \\
      \hline
      Deployment Refinement   
      & 12.92 &  8.32 &  20.84 & 11.31 & 15.70 &  20.82 &  20.82 &  20.83 &  5.61 \\
      \hline
    \end{tabular}%
  }
\end{table}

\section{Tables of Empirical Comparison to Existing Methods}\label{app:empirical-comarison-tables}

\paragraph{Comparison to audit2rbac}
Table~\ref{tab:audit2rbac-empricial-comparison} presents a comparison of Role Creation performance between audit2rbac~\cite{liggitt_audit2rbac_2023} and \name.
\name's pod-centric approach achieved perfect scores, while user-centric audit2rbac overgenerated rules, and despite full recall, yielded low precision due to excessive FPs.

\begin{table}[h]
\centering
\caption{Comparison of Role Creation performance between audit2rbac~\cite{liggitt_audit2rbac_2023} and \name (using GPT-4o).}
\label{tab:audit2rbac-empricial-comparison}
\resizebox{\columnwidth}{!}{
\begin{tabular}{|l|l|l|l|l}
\hline
\textbf{Method} & 
\textbf{Precision} & \textbf{Recall} & \textbf{F1-Score}  \\ \hline
audit2rbac
& $0.208 \pm 0.042$ & $1.00 \pm 0.0$ & $0.343 \pm 0.057$  \\ \hline
\name
& $1.00 \pm 0.0$ & $1.00 \pm 0.0$ & $1.00 \pm 0.0$ \\ \hline
\end{tabular}
}
\end{table}

\paragraph{Comparison to KUBETEUS}
We evaluated KUBETEUS~\cite{kimkubeteus} and \name on NetworkPolicy Creation using the same SLM, Deepseek-Coder-1.3B~\cite{guo2024deepseekcoderlargelanguagemodel}, to ensure a fair comparison under constrained compute environments. 
As shown in Table~\ref{tab:kubeteus-empricial-comparison}, despite modest absolute performance due to the SLM’s limited capacity, \name consistently generated more secure NetworkPolicies, with higher scores across all metrics.

\begin{table}[h]
\centering
\caption{Comparison of NetworkPolicy Creation performance between KUBETEUS~\cite{kimkubeteus} and \name, both using the Deepseek‑Coder‑1.3B SLM.}
\label{tab:kubeteus-empricial-comparison}
\resizebox{\columnwidth}{!}{
\begin{tabular}{|l|l|l|l|}
\hline
\textbf{Method} & 
\textbf{Precision} & 
\textbf{Recall} & 
\textbf{F1-Score} \\ \hline
KUBETEUS
    & $0.360 \pm 0.339$ & $0.238 \pm 0.248$ & $0.275 \pm 0.270$  \\ \hline
\name
    & $0.444 \pm 0.355$ & $0.265 \pm 0.234$ & $0.332 \pm 0.250$ \\ \hline
\end{tabular}
}
\end{table}

\section{Attack Surface Reduction Prompt Chains Illustrations}

This section includes illustrations of two attack surface reduction prompt chains (Sec.~\ref{subsec:Attack_Surface_Reduction_Prompt_Chains}): NetworkPolicy Creation and Deployment Refinement.

\paragraph{NetworkPolicy Creation Prompt Chain}\label{app:networkpolicy_creation_prompt_chain}

\begin{figure*}[h]
    \centering
    \includegraphics[width=1.35\textwidth, angle=90]{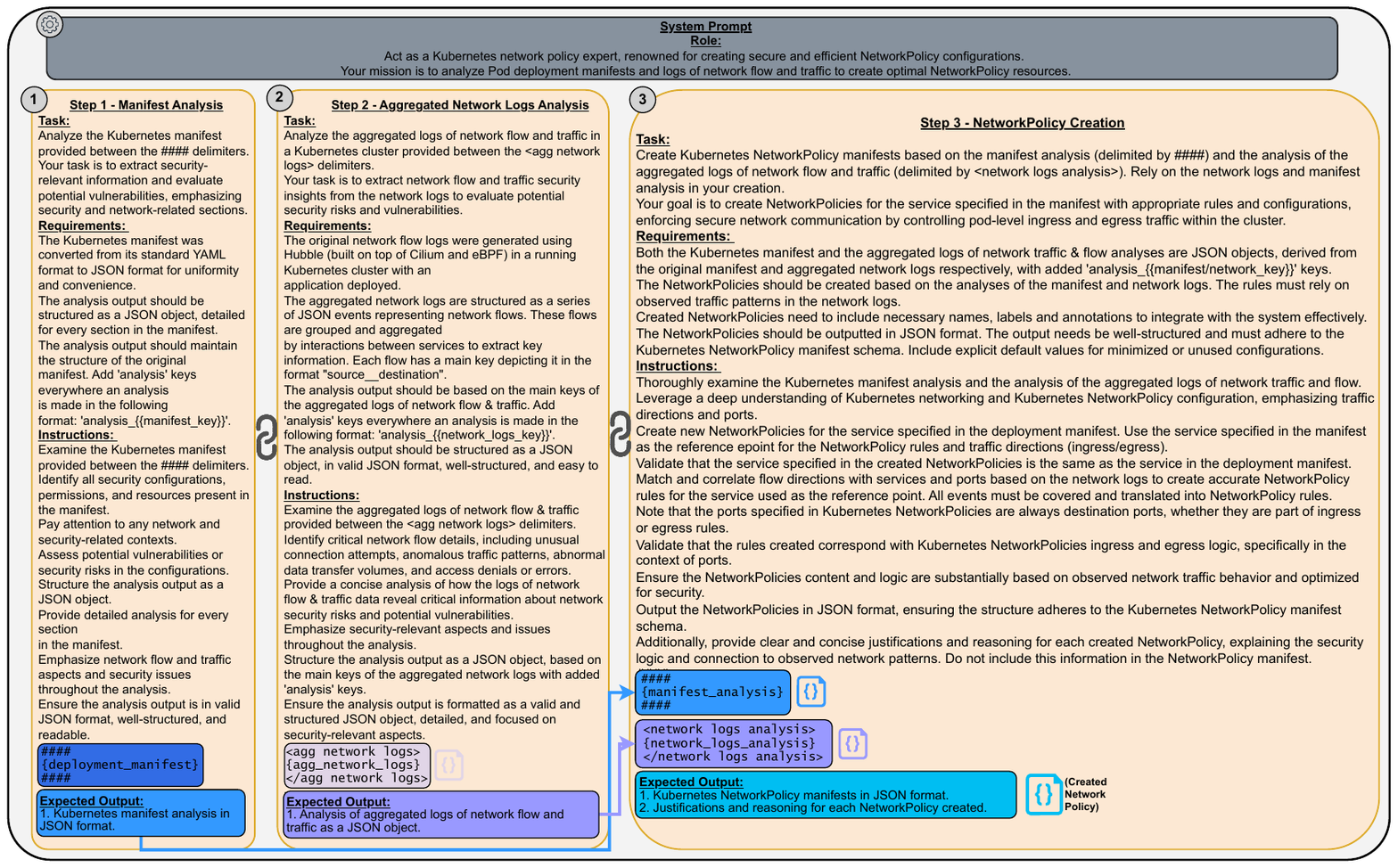}
    \caption{NetworkPolicy Creation prompt chain.}
    \label{fig:networkpolicy_creation_prompt_chain}
\end{figure*}

Fig.~\ref{fig:networkpolicy_creation_prompt_chain} presents the complete prompt chain used for creating K8s NetworkPolicies based on Deployment manifests and aggregated network logs (ANLs).
The process begins by extracting relevant network configurations from Deployment manifests, followed by analyzing ANLs to identify communication patterns, detect anomalies, and flag unauthorized access attempts. 
The prompt chain then guides the LLM through the generation of fine-grained ingress and egress rules that enforce secure, least-privilege inter-Pod communication. 
Each step builds on the previous one to ensure coherent and progressive refinement, ultimately producing NetworkPolicy manifests that align with observed runtime behavior and mitigate potential vulnerabilities.

\paragraph{Deployment Refinement Prompt Chain}\label{app:deployment_refinement_prompt_chain}

\begin{figure*}[h]
    \centering
    \includegraphics[width=\textwidth, trim=20 10 20 10, clip]{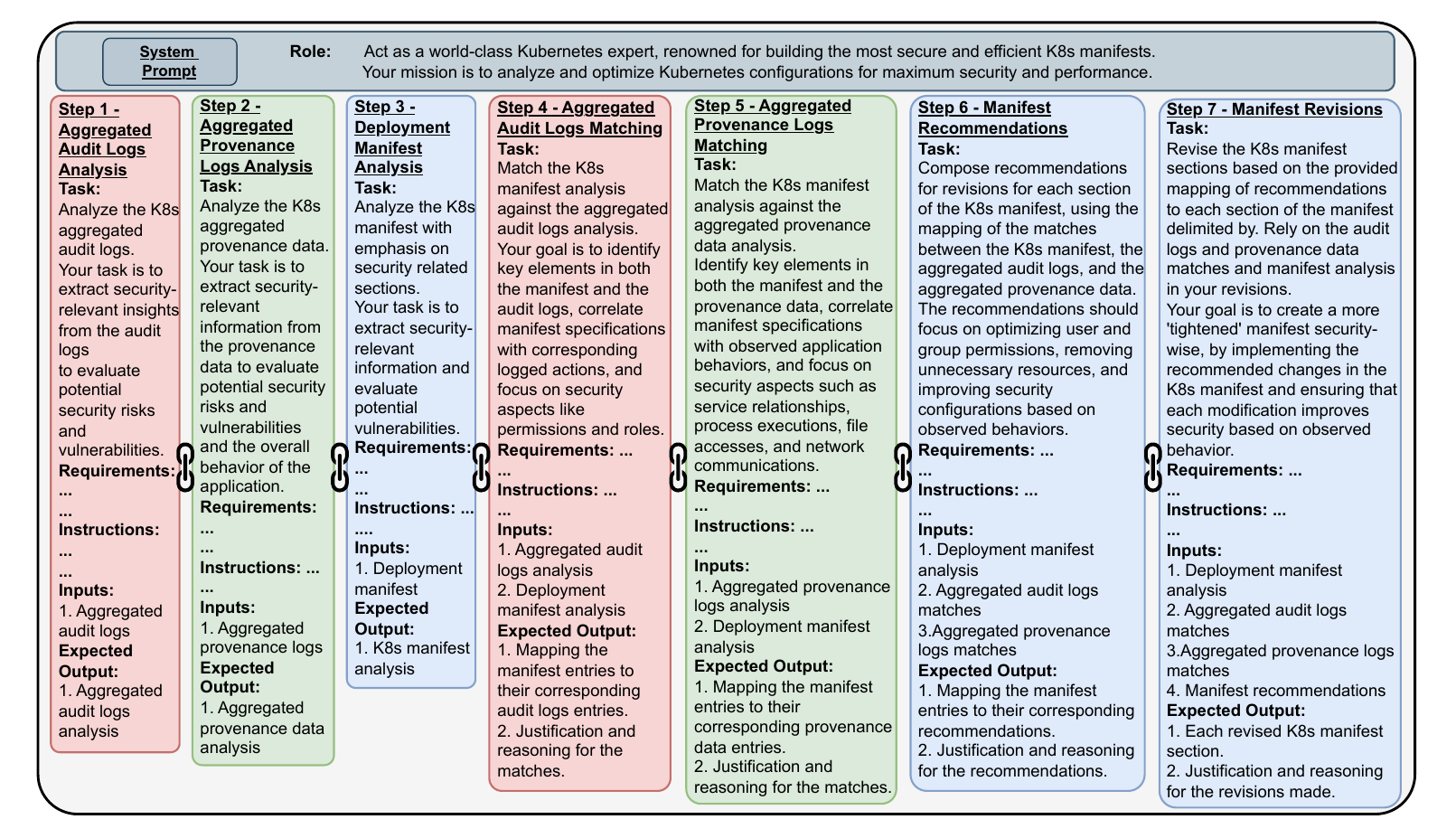}
    \caption{High-level illustration of the Deployment Refinement prompt chain.}
    \label{fig:high_level_deployment_refinement}
\end{figure*}

Fig.~\ref{fig:high_level_deployment_refinement} presents a high-level overview of the Deployment Refinement prompt chain, which guides the LLM through a structured, multi-step analysis to harden Kubernetes Deployment manifests. 
The process begins with the extraction of security-relevant information from aggregated audit logs (Step 1) and aggregated provenance logs (Step 2), both aimed at identifying potential vulnerabilities and deviations in run-time behavior. 
The Deployment manifest is then analyzed (Step 3) with a focus on security-critical sections such as permissions and role specifications.
Subsequent steps involve matching the manifest specifications against the audit log analysis (Step 4) and provenance log analysis (Step 5) to establish precise correlations between declared configurations and observed behaviors. 
This matching step ensures that recommendations (Step 6) are grounded in concrete run-time evidence, targeting the removal of unnecessary resources, over-privileged permissions, and redundant configurations. 
The final stage (Step 7) revises the manifest by implementing the validated recommendations, producing a hardened, security-optimized version aligned with least-privilege and minimal-attack-surface principles.

\end{document}